Displacement and disconnection: the impact of violence on migration networks and highway traffic in Mexico

Michele Coscia and Roxana Gutiérrez-Romero[♣]

**Abstract**

We examine how violence affects migration flows and, crucially, how it reshapes the strength of migration networks. This strength is measured by the intensity of migration between areas, taking into account that some routes become more prominent while others fade over time, an aspect often overlooked by traditional studies. Using a novel network algorithm and Mexican census data from 2005 to 2020, we first quantify changes in the strength of domestic and international migration networks across all Mexican municipalities. We exploit variation in local homicide rates, using exogenous fuel price increases and municipalities' proximity to oil pipelines as instruments, to estimate the causal impact of violence on migration. During our study period, following intensified government crackdowns on drug trafficking organizations, many criminal groups fragmented and turned toward large-scale oil theft, driving sharp increases in violence in areas with oil pipelines, particularly when fuel prices rose. The findings show that rising violence increased emigration flows, predominantly within Mexico, and strengthened the intensity of emigration networks both domestically and toward the United States. Although violent municipalities continued to receive new residents, the rise in emigration was larger. Increasing homicide rates led to at least an additional 1.12 million people emigrating domestically and 50,200 fewer Mexicans returning from the United States. Violence also eroded regional connectivity, causing a long-term decline in daily vehicle traffic on highways linking violent areas to the rest of the country.

---

[♣] We acknowledge funding from the Global Challenges Research Fund (GCRF) [RE-CL-2021-01]. We are also grateful to Constantino Carreto, Daniel Chavez, Tania Rodríguez and Nayely Iturbe for outstanding research assistance.
Corresponding authors: Michele Coscia, IT University of Copenhagen, mcos@itu.dk.
Roxana Gutiérrez-Romero, Queen Mary University of London, r.gutierrez@qmul.ac.uk.

## 1. Introduction

Violence, ranging from armed conflicts to organized crime, can influence migration patterns in complex and unpredictable ways. While violence typically increases the risk of victimization and can drive people away (Zolberg, Suhrke, and Aguayo 1993; Blattman and Miguel 2010), strong local ties and economic opportunities may anchor residents or even attract newcomers (Adhikari 2013). Mexico presents a compelling case study in understanding the multifaceted relationship between violence and migration. Since 2006 the government's offensive against drug cartels has fragmented these organizations, prompting them to diversify into other illicit activities such as large-scale oil theft. This shift intensified violence and insecurity, contributing to over 300,000 homicides linked to organized crime thus far (Calderón et al. 2015; Battiston et al. 2024). The escalation of violence has also had broad social and economic consequences, including setbacks in education, higher unemployment, increased political violence, and greater fear of victimization (Michaelsen and Salardi 2020; Dell 2015; Gutiérrez-Romero 2016; Velásquez 2019; Magaloni et al. 2019). Despite these multiple negative outcomes, the impact of violence on migration remains a subject of ongoing debate. Anecdotal reports suggest that violence has displaced thousands of families, yet economic constraints have prevented others from leaving (Basu and Pearlman 2017). Paradoxically, some violent regions, particularly urban centers, have continued to attract immigrants (Atuesta and Paredes 2015), obscuring the overall net effect of violence on migration.

This paper examines two central questions: to what extent has violence driven internal displacement within Mexico, and how has it affected international migration, particularly to the United States? We move beyond conventional analyses that focus exclusively on how socioeconomic push and pull factors influence the volume of migration flows. We show that these factors, including violence, can also reshape the strength of both domestic and international migration networks in ways that aggregate flow measures may hide. External shocks such as sudden rises in homicide rates may leave overall migration numbers unchanged but can significantly reconfigure where people are moving from and to, with longstanding migration routes breaking down and new ones emerging in response. To estimate the causal effect of violence on these network shifts, we rely on an instrumental variables strategy that exploits plausibly exogenous variation in local homicide rates.

We begin by quantifying structural changes in Mexico's migration networks using individual-level census data from 2005 to 2020 and the Noise-Corrected (NC) algorithm, originally proposed by Coscia and Neffke (2017). This algorithm allows us to analyze more than three million potential bilateral connections across Mexico's 2,454 municipalities, including their international links. Traditional network methods often overlook noise and measurement error in migration data. In contrast, the NC algorithm models the expected strength of a connection and compares it to observed flows, enabling us to detect whether networks have expanded or contracted, even when initial links are weak or absent. Crucially, the algorithm summarizes structural changes (across emigration and immigration, domestic and international flows) into a single score for each municipality, without using or identifying any socioeconomic variables that may be driving those shifts.

We then examine the key factors behind these network changes, focusing on the role of homicide rates as a proxy for violence. A central challenge is endogeneity: both violence and migration may respond to unobserved local conditions, such as institutional weakness or socio-economic and cultural context. To address this concern, we use instrumental variables. Our identification strategy exploits a major shift in criminal activity during our study period, Mexico's drug war and the rise of large-scale fuel theft. As the federal government increased reliance on fuel imports and cut fuel subsidies, gasoline prices rose substantially, making stolen fuel highly profitable, on par with drug trafficking (Ferri 2019). In response, criminal organizations began a deadly war targeting municipalities located near oil pipelines, which offered access points for theft. This deadly competition triggered sharp increases in local homicide rates along the fixed pipeline network.

We instrument for changes in municipal homicide rates using variation in real fuel prices and its interaction with the inverse distance from each municipality's centroid to the nearest oil pipeline. While these factors jointly affect the returns to fuel theft, and therefore generate exogenous variation in violence, they are unlikely to directly influence migration decisions. Fuel prices are set by national energy policy and reflect global market trends, while the pipeline infrastructure was installed decades before the drug war, is buried underground, and has no direct economic relevance to households beyond its role in facilitating violence. First-stage estimates confirm that the instruments strongly predict changes in homicide rates and reveal substantial endogeneity in the violence–migration relationship.

Our paper makes several important contributions. First, using network analysis and instrumental variables, we advance research on migration by providing causal evidence on how local violence reshapes internal and international migration flows (e.g. Adhikari 2013; Basu and Pearlman 2017; Orozco-Aleman and Gonzalez-Lozano 2018). Escalating homicide rates significantly increased internal displacement within Mexico between 2005 and 2020. Despite municipalities with rising violence receiving approximately 941,000 new domestic immigrants, the overall outcome was a substantial net outflow of residents. We estimate that increasing violence drove over 1.12 million individuals, around 6% of total internal migration, to relocate to other municipalities. This pattern highlights the complex role that violence plays in shaping internal migration dynamics.

Second, our paper brings a network perspective to the literature, showing how violence has impacted the origin-destination structure of domestic and international migration networks across millions of links, an aspect that traditional net flow analyses cannot detect (Coscia and Neffke 2017). Specifically, emigration networks expanded both domestically and toward the U.S., reflecting broader shifts in migratory routes triggered by rising homicide rates. Data from both the U.S. and Mexico since 2008 suggest that more Mexican migrants have returned from the U.S. than have left for it (Gonzalez-Barrera 2015). This period has also been characterized by an increase in the return migration of Mexicans from the U.S. (Pearson 2021; Rosenblum et al. 2014). While this broader trend is consistent with our findings, our analysis reveals a more nuanced impact of violence on these migration dynamics. We show that rising homicide rates have not only deterred return migration but have also spurred additional emigration to the U.S. Between 2005 and 2020, approximately 22,399 individuals emigrated to the U.S. as a direct consequence of increased violence, while about 50,249 Mexicans were effectively discouraged from returning for the same reason. These impacts account for around 1% of total emigration to the U.S. and a 3% reduction in return migration flows from the U.S. Our network analysis further reveals that although the structure of return migration networks remained stable, their overall volume declined, while emigration networks toward the U.S. expanded. These findings challenge conventional views by highlighting the dual role of violence in simultaneously suppressing return migration and driving emigration flows, predominantly within Mexico.

Our network-based results also help reconcile mixed findings in the existing literature. For instance, earlier studies in Mexico have shown that drug-related homicides pushed individuals to relocate to less violent states, even sacrificing wages (Atuesta and Paredes 2015). This perspective, nonetheless, overlooks that migration occurs predominantly within states. At the municipal level, findings on net migration numbers have been mixed: while some suggest that lower-violence areas attract migrants (Gutiérrez-Romero and Oviedo 2018), others find no impact (Basu and Pearlman 2017). A key reason for these discrepancies is that previous studies have not accounted for the origin-destination structure of migration flows. Even if net migration numbers appear stable, significant shifts in the network dynamics may still be occurring, reflecting deeper changes in how people move between locations. Studies of U.S.-bound migration have been similarly inconsistent: while some link higher homicide rates in border cities to more emigration (Orozco-Aleman and Gonzalez-Lozano 2018; Rios Contreras 2014; Daniele, Le Moglie, and Masera 2023), others find that municipalities with more violence sent fewer people, likely due to the high migration costs (Basu and Pearlman 2017). By analyzing millions of potential migration connections across Mexican municipalities and to the U.S., our network approach shows that violence not only displaced individuals but also reshaped and expanded emigration routes, capturing patterns that would remain invisible in net flow statistics.

Third, we contribute to the literature on migration by showing that violence not only displaces populations but also can disrupt the long-term integration of violent municipalities with the rest of the country (Orozco-Aleman and Gonzalez-Lozano 2018; Gutiérrez-Romero 2024). To examine this broader effect, we extend our network analysis to daily vehicle traffic across the national highway system connecting all Mexican municipalities. We focus on changes between 2005 and 2015, a period that predates the likely disruption in mobility caused by the COVID-19 lockdowns. If violence reduces the attractiveness of an area for residence or commerce, this should be reflected in declining vehicle flows linking that area to the rest of the country.

Because changes in vehicle flows are likely to respond directly to fuel prices, we adopt a different identification strategy to estimate the effect of violence on highway connectivity. Specifically, we instrument for changes in municipal homicide rates using changes in the number of drug trafficking organizations (DTOs) operating across municipalities. Changes in DTO presence increase local homicide rates but are not expected to influence highway traffic directly, except by raising violence and the associated risk of victimization. The estimates show that between 2005 and 2015, national traffic volumes increased by 154% over this period. However, violence substantially reduced intermunicipal mobility. Municipalities with rising homicide rates lost nearly 110 million daily vehicle trips connecting them to the rest of the country, equivalent to 49% of all intermunicipal daily highway traffic recorded in 2015. These losses reflect reductions in long-distance travel connecting municipalities rather than local circulation, pointing to a broader erosion of regional connectivity.

The remainder of the paper is structured as follows. Sections 2 and 3 provide an overview of data and background. Section 4 explains the NC algorithm. Section 5 discusses the empirical strategy and results. Section 6 shows the robustness check of the changes in the highway network. Section 7 concludes.

## 2. Data

### *2.1 Migration data*

To examine changes in municipalities' migration networks and flows, we use the 2010 and 2020 population censuses and the 2015 intercensal survey, all collected by Mexico's National Institute of Statistics and Geography (INEGI) and made publicly available as microdata samples. These include sampling weights to ensure representativeness at both the municipal and national levels.

Each source captures migration that occurred during the five years preceding the interview date. That is, the 2010 census covers migration moves from 2005 to 2010, the 2015 survey from 2010 to 2015, and the 2020 census from 2015 to 2020. Notably, the 2020 census was not affected by the COVID-19 pandemic, as fieldwork concluded shortly before the onset of lockdowns.

The census captures migration history by asking respondents where they lived five years before the interview. Those who lived in Mexico report their municipality and state; those who lived abroad report their country. The census does not record moves within the same municipality, the exact timing of moves across municipalities, or whether a person moved multiple times. However, by comparing respondents' current residence and their reported residence five years earlier, both stated in the same interview, we can identify whether they moved during the five years preceding the interview. Because the census also records each person's nationality, we can identify Mexican nationals who returned from abroad. Among these return migrants, we distinguish those who came back from the U.S. and from those who returned from other countries.

### *2.2 Identifying changes in migration flows and networks across censuses*

As explained in Section 4, we study changes in migration networks by comparing their evolution across consecutive census waves. That is, we quantify how migration networks have changed over time by comparing the 2010 census to the 2015 intercensal survey, and the 2015 survey to the 2020 census. Because each census captures migration flows over the preceding five years, this yields two periods of analysis: 2005–2010 and 2010–2015 for the first comparison, and 2010–2015 and 2015–2020 for the second.

During our period of analysis, only the 2010 and 2020 censuses asked respondents to indicate whether any household member had emigrated abroad. The dates of international emigration are recorded with greater precision than those of domestic migration. To maintain consistency with the periods we analyze for domestic migration analysis, we capture international emigration that occurred five years before the interview, as well as in the year 2010. This allows us to compare changes in international emigration networks for the years 2005 and 2010, as well as 2010 and 2015.

*2.3 Homicide rate and additional variables*

A central variable in our analysis is the municipal homicide rate, which we use as a proxy for local violence. Some earlier studies on migration and Mexico's drug war have focused on drug-related homicides during 2006-2012 (Dell 2015; Gutiérrez-Romero and Oviedo 2018; Atuesta and Paredes 2015), data released by the National Council of Public Security. We instead rely on overall homicide rates, which capture a broader measure of violence, albeit one largely driven by the ongoing conflict between organized criminal groups. This choice is partly practical: the Mexican government discontinued the publication of drug-related homicide data after 2012 due to political sensitivities and classification challenges. Nonetheless, total homicides provide a more consistent and comprehensive indicator of the risk of victimization and generalized violence over time. We construct municipal-level homicide rates per 100,000 inhabitants using death records from INEGI, which draws primarily on data from Forensic Medical Services and Civil Registry Offices, with supplementary information from Public Prosecutor's Offices. Population figures are based on estimates from CONAPO. To assess how fluctuations in homicide rates influence shifts in migration patterns, we estimate the changes in these rates between the years 2005 and 2010, and between 2010 and 2015.

We incorporate three additional controls, all of which are measured at the municipal level. First, we calculate the change in poverty rates at quinquennial intervals from 2005 to 2015 using census data. We define poverty based on a per capita income threshold of five dollars per day.[1] The poverty rate in 2005 cannot be approximated directly from the intercensal survey because respondents were not asked about their household income. To address this, we estimate the 2005 poverty rate based on census data from 2000. Given the continually high levels of poverty over this period, this is a good proxy for the 2005 levels.

Second, we estimate the changes in inequality, measured by the Gini coefficient for five-year change quinquennial points between 2005 and 2015. This coefficient is based on

[1] Consejo Nacional de Evaluación de la Política de Desarrollo Social (CONEVAL) has produced various poverty statistics for the country. These measures have changed over time and are not directly comparable across our period of analysis. Thus, we adopted the alternative poverty line of five dollars per day per capita to ensure consistency in our assessment.

CONEVAL's estimates from the *Encuesta Nacional de Ingresos y Gastos de los Hogares* (ENIGH). Third, we estimate changes in annual nighttime light-per-capita, aggregated over the periods 2001-2005, 2006-2010, and 2011-2015, as a proxy for wealth. These data were obtained from the Earth Observation Group, Payne Institute for Public Policy (Elvidge et al. 2017).

In the robustness section, we assess how violence impacts vehicle traffic along highways. To this end, we use the change in the annualized daily average number of vehicles passing through municipalities' highway networks, which connect these areas to the rest of the country. We use vehicle flow data for 2005 and 2015 only. This allows us to capture conditions before and after the onset of the drug war while avoiding the 2020 data, which was likely affected by pandemic-related lockdowns. The vehicle traffic data are sourced from the Ministry of Communications and Transport.

## 3. Background: Dynamics of violence and migration in Mexico

When President Felipe Calderón assumed office in Mexico in 2006, the country had one of the lowest homicide rates in Latin America, with 10 homicides per 10,000 inhabitants (Wainwright 2017). Drug-related homicides were nonetheless escalating in northern cities like Tamaulipas, Chihuahua, and Sinaloa, as well as in the port city of Michoacán, a key entry point for drugs from South America. To tackle this violence and regain credibility after winning a closely contested presidential election, he launched a large-scale military offensive. Federal troops were deployed to combat the drug cartels, marking the beginning of the 'war on drugs' strategy, which led to a dramatic increase in violence. The homicide rate more than doubled by 2011. Homicides slightly dropped temporarily toward the end of Calderón's term in 2012 as the security focus shifted to dismantling the Gulf and Zetas cartels' logistical networks in high-violence states. Violence escalated again under President Enrique Peña Nieto, as more drug lords were arrested or killed than ever before.[2] Homicides peaked in 2018 and gradually stabilized under President Andrés Manuel López Obrador (AMLO) but remain at historically high levels (Fig. 1).

---

[2] During Calderón's six-year presidency, 25 drug lords were apprehended. In contrast, 96 drug lords were apprehended and 14 were killed during Peña Nieto's presidency, with the majority (84%) occurring during his first two years in office, resulting in more succession disputes for leadership and territory (Dittmar 2018).

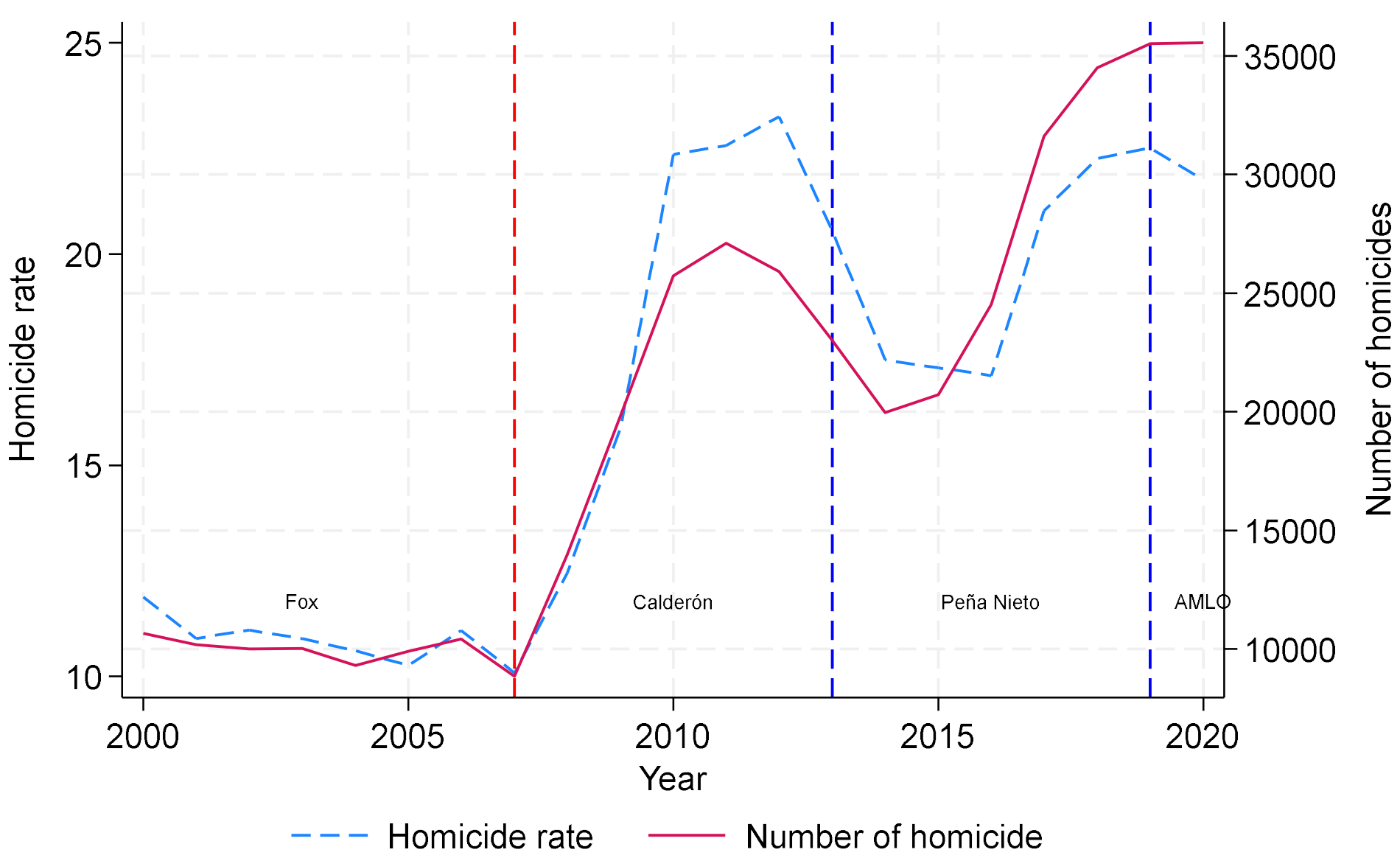

**Fig. 1.** Number of homicides and homicide rates per 100,000 inhabitants in Mexico

Government crackdowns on drug trafficking prompted criminal organizations to diversify into new criminal sectors, pushing DTOs into large-scale oil theft (Battiston et al. 2024). By 2010, major DTOs like the Zetas and Gulf Cartel began capitalizing on gasoline theft. This process includes bribing municipal mayors and local police to enable large-scale oil extraction, storage in trucks, and widespread distribution of stolen fuel on the black market. Mayors, who also head local police, have become targets of DTOs when they resist such pressure, fueling ongoing political violence against political candidates and mayors in municipalities near oil pipelines (Gutiérrez-Romero and Iturbe 2024). Employees of Petróleos Mexicanos (Pemex) are also key collaborators for DTOs, providing access to pipeline information and facilitating the extraction of gasoline and diesel from the oil pipeline network. Over 90% of stolen fuel is sold through regular petrol stations at standard market prices, as well as a few transport businesses, with a portion even smuggled into the U.S. (Stevenson 2017).

The deregulation of gasoline prices, combined with the high profitability of gasoline theft, further incentivized the illegal oil market. This deregulation was driven by Mexico's growing reliance on imported gasoline to meet domestic demand as Pemex, once a cornerstone of the national economy, faced declining productivity due to aging infrastructure and insufficient investment in exploration. The Calderón and Peña Nieto administrations responded by introducing sudden and unexpected price hikes, known as *gasolinazos*, to reduce oil subsidies and align domestic oil prices with global markets. This process culminated in a 20% price increase in 2017 and the deregulation of gasoline prices in 2018. Gasoline prices in the country have remained close to historical highs, driven by international prices, the further elimination of all gasoline and diesel subsidies.[3]

Notably, the Cartel Jalisco Nueva Generación (CJNG), which expanded into oil theft by the mid-2010s, and the Santa Rosa de Lima Cartel, which emerged in 2017 with a focus on oil theft, intensified the violence in regions where they competed for control (International Crisis Group 2022). In 2018, President Andrés Manuel López Obrador aimed to combat oil theft by temporarily using tanker trucks for gasoline distribution instead of the pipeline network (Jones and Sullivan 2019). Although this measure initially reduced oil theft, the problem resurfaced and spread, resulting in an estimated $5.2 billion in oil theft by 2023, despite stricter penalties (Arzate 2023).

Fig. 2 illustrates the sharp increase in the number of illicit pipeline taps detected by the government, along with gasoline and diesel prices during 2000-2020. These pipelines require highly specialized equipment to be tapped, making it highly likely that they were primarily targeted for oil theft. The number of illicit pipeline taps increased from 200 in 2006 to over 15,000 in 2019, reflecting how the rapid growth of oil theft coincided with both rising homicide rates, diesel and gasoline prices. Diesel is primarily used in the commercial and transportation sectors. Magna has an octane rating of 87, which is suitable for standard engines, while premium, with an octane rating of 91 or higher, meets international gasoline standards and provides better performance for modern engines. Despite these differences, all types of gasoline and diesel rose in price.

[3] Although gasoline prices were deregulated in 2018, magna, premium gasoline, and diesel retained partial subsidies for some time. By 2022, the subsidies for premium gasoline and diesel were fully removed, while magna gasoline continued to receive a small subsidy (0.78 to 0.26 pesos per liter) until September 2024, when it was eliminated entirely (Proceso 2024).

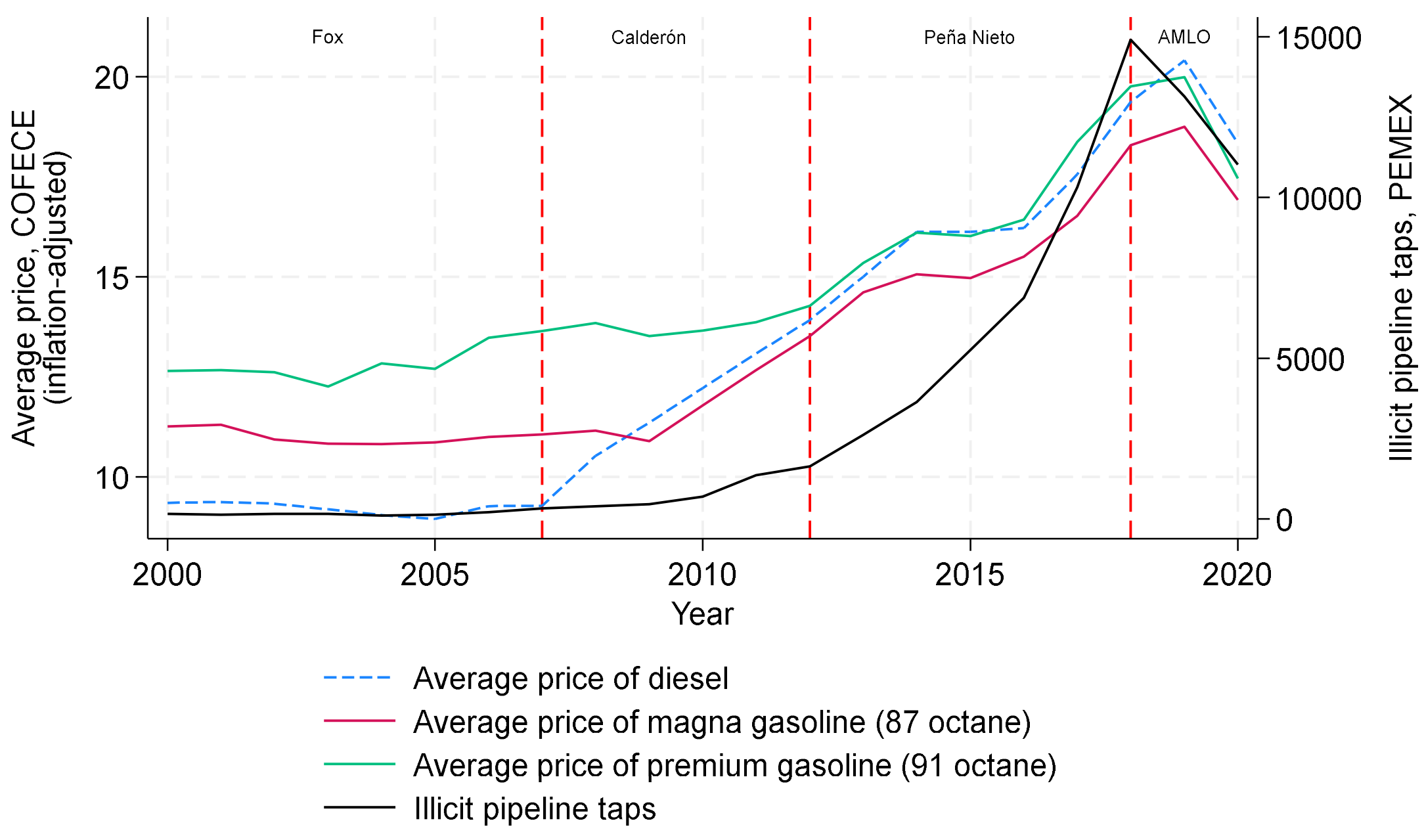

**Fig. 2.** Average price of gasoline in pesos, and number of illicit pipeline taps

Fig. 3 shows that between 2005 and 2010, homicide rates spiked in northern states like Chihuahua and Tamaulipas and along the coasts, as criminal organizations fought for control of drug routes and oil infrastructure used for large-scale theft. By the second phase of the drug war, between 2010 and 2015, violence had pushed into central Mexico, reflecting the growing importance of large-scale oil theft.  Fig. 4 maps the network of pipelines transporting crude oil and refined products, highlighting the overlap between contested infrastructure and areas of escalating violence.

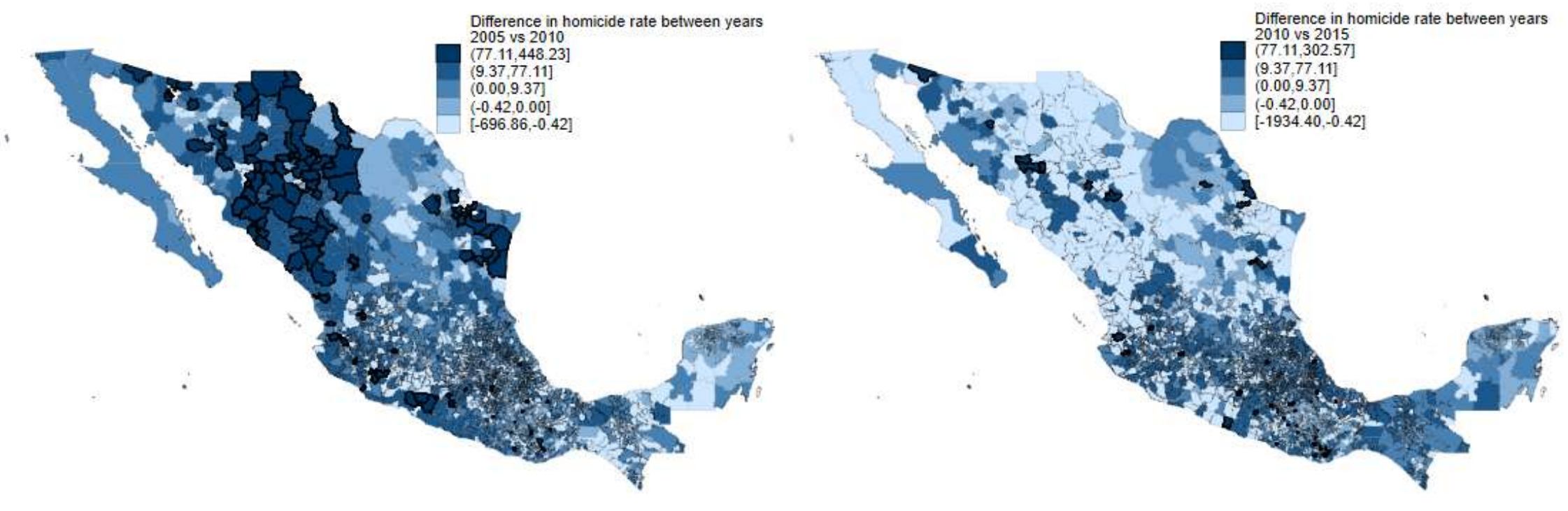

**Fig. 3.** Changes in homicide rates

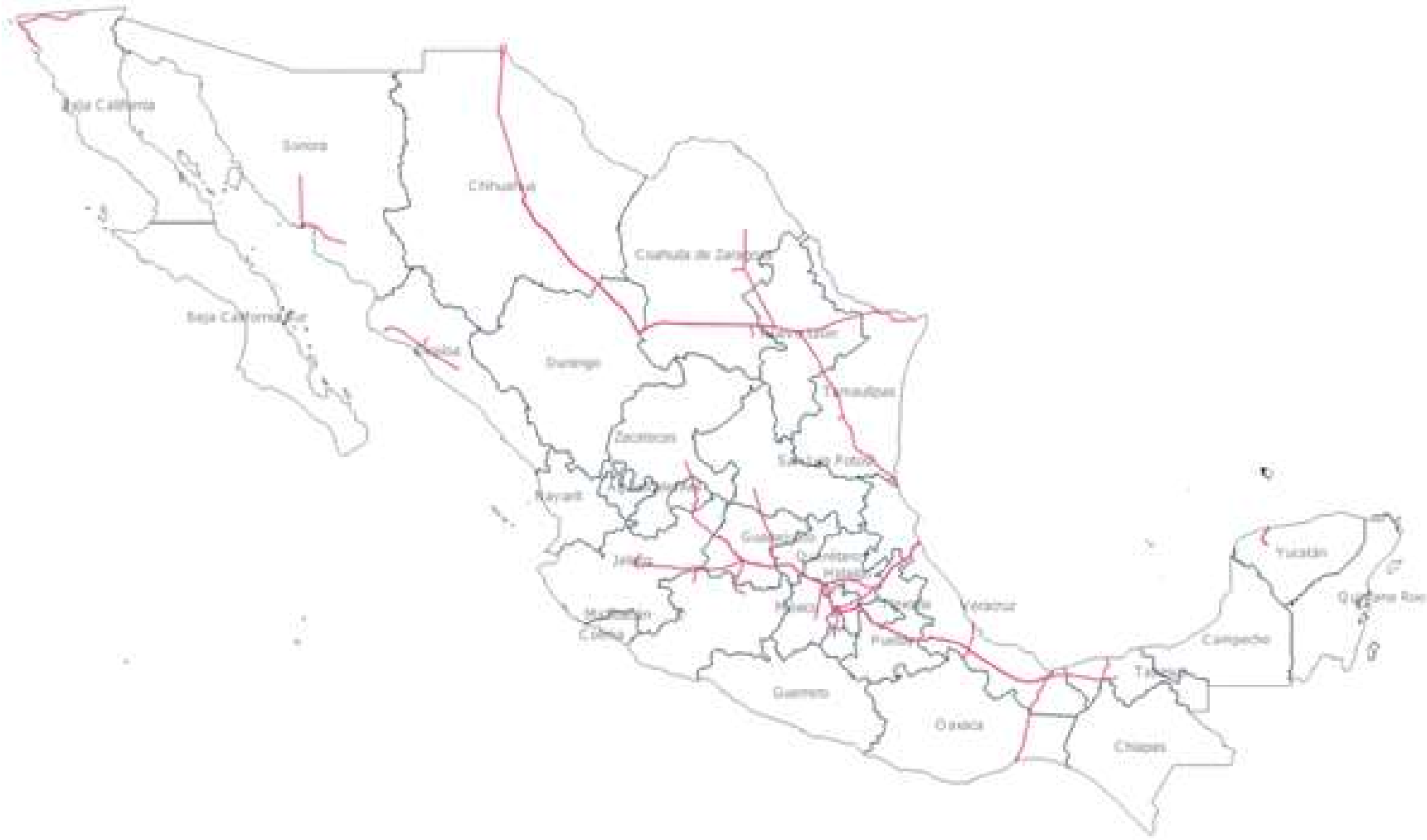

**Fig. 4.** Oil pipeline network

Domestic migration seemed to have reacted to the growing violence. Comparing the migration flows captured by the 2010 census and the 2015 intercensal survey, northern and central regions experienced both high number of domestic emigrants and immigrants (Fig. 5). Comparing instead the 2015 survey and the 2020 census, emigration surged in violence-stricken areas, while northern border states continued to attract migrants. In violence-prone regions, emigration to the U.S. initially declined but later rebounded, while return migration remained low throughout (Fig. 6).

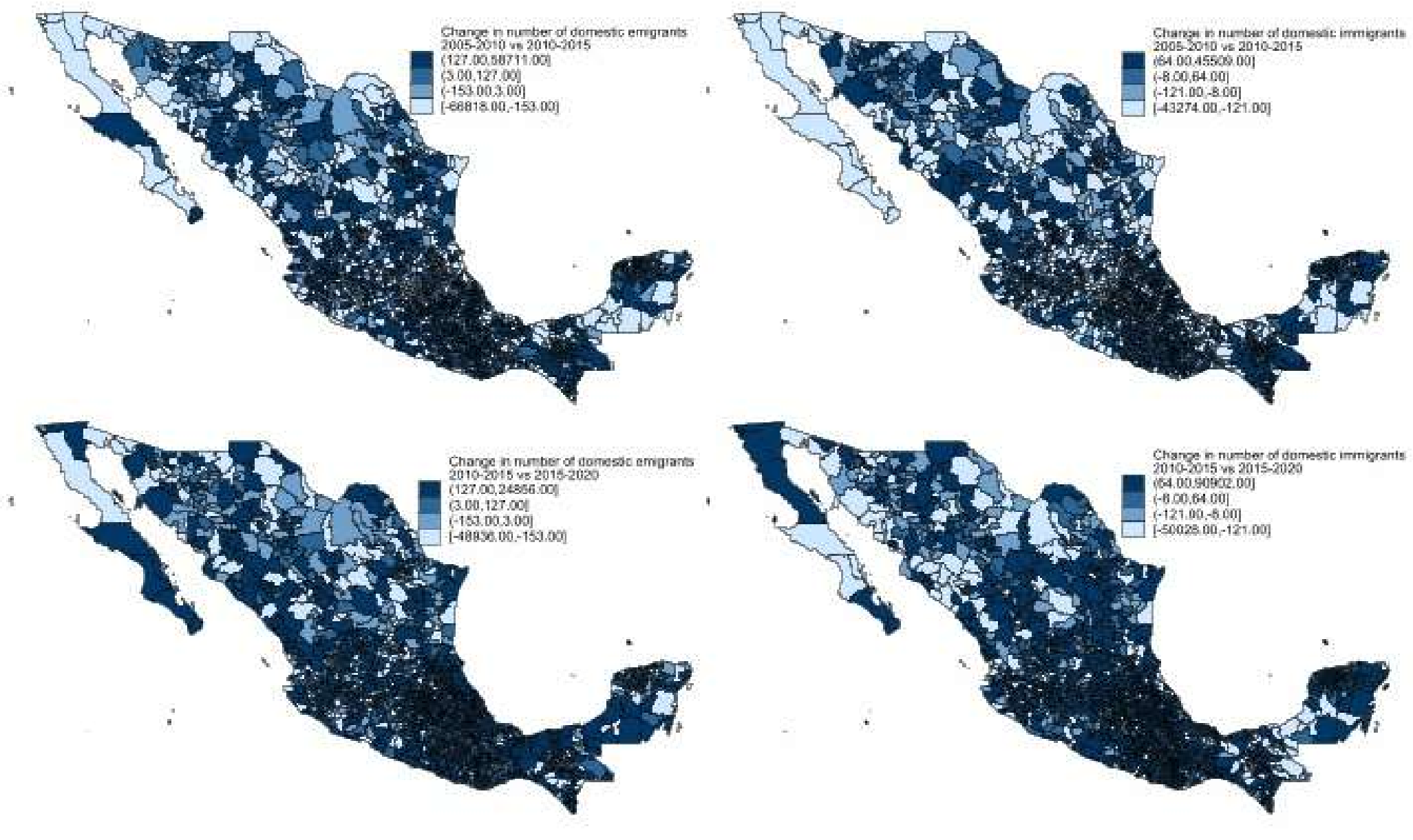

**Fig. 5.** Change in the number of domestic migrants across municipalities

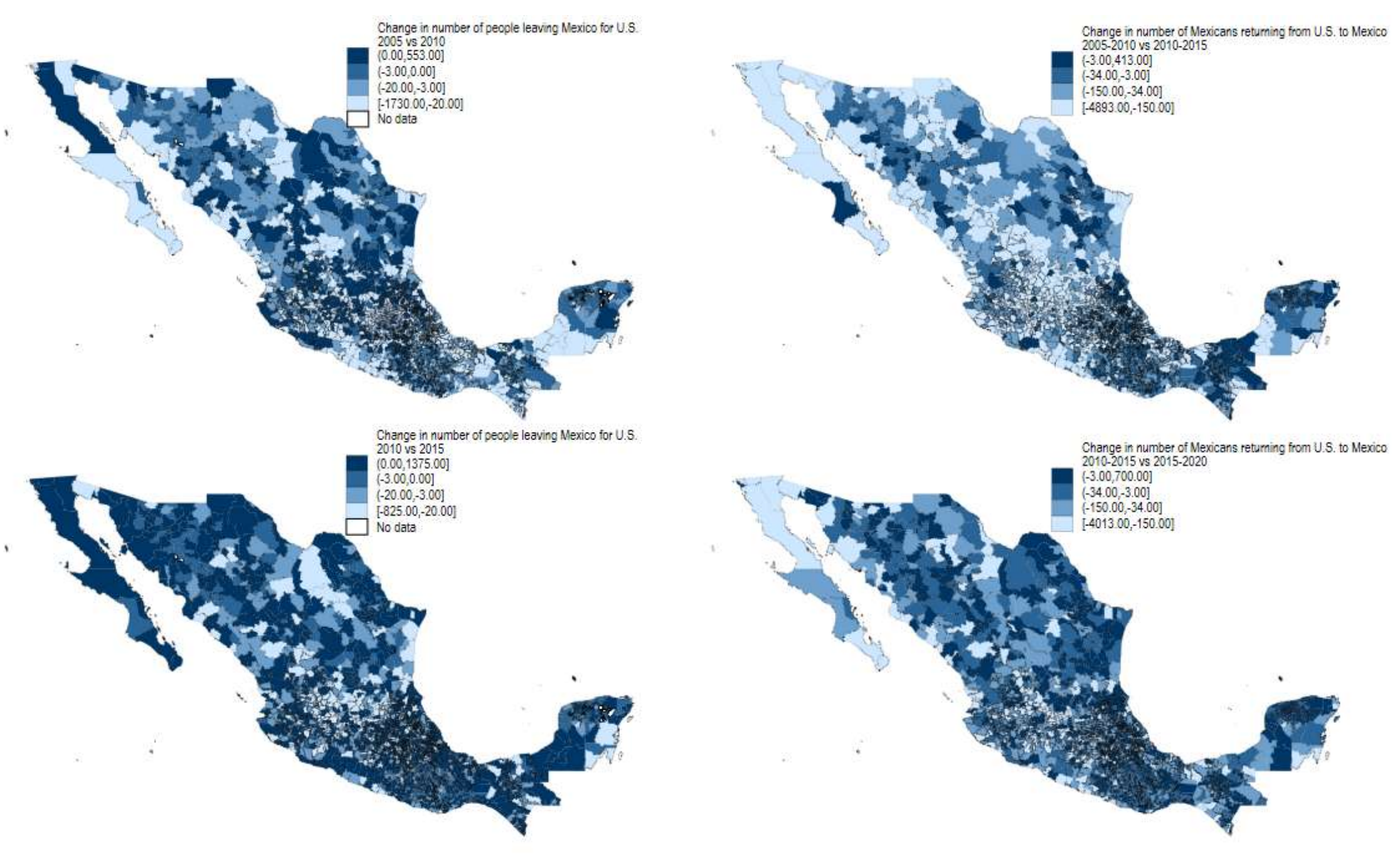

**Fig. 6.** Change in the number of people that left for the U.S. and Mexicans that returned from the U.S.

As previously mentioned, changes in the number of migrants do not fully reveal the underlying changes in the migration networks that each municipality may be experiencing. To identify these structural changes, we apply the Noise-Corrected backbone algorithm, which we describe next.

### 4. Network analysis

We use network analysis to track changes in domestic and international migration across each of Mexico's 2,454 municipalities. The network-based approach takes into account the origin and destination of migration flows for each area. This allows us to measure whether a municipality's connections with other areas, both domestically and internationally, are becoming stronger or weaker over time.

Popular network analysis methods, such as the Disparity Filter (DF), often assume that if no migration connection is recorded between two areas, it indicates a complete absence of migration (Serrano, Boguñá, and Vespignani 2009). This approach fails to account for potential measurement errors or data noise, which can lead to biased network representations.

To overcome this limitation, we apply the Noise-Corrected (NC) backbone algorithm developed by Coscia and Neffke (2017). The NC algorithm addresses these issues by accounting for noise in the data and providing a more accurate calculation of changes in both the structure and strength of migration networks using a probabilistic approach. Importantly, this approach solely focuses on identifying the network structure itself and its changes, without incorporating or needing socioeconomic factors that may drive these changes.

The NC algorithm reveals the migration network and changes in such a network over time in three main steps. First, it compares the observed migration flows between municipalities, known as edge weights, to what would be expected under a null model. The null model assumes that the likelihood of migration flows between any two areas is proportional to the total number of migrants leaving one area and entering another. The expected number of interactions between two areas, i.e., node pair ($i$, $j$) is expressed as in equation (1). This formula provides the baseline expectation for each migration connection, the expected edge weight, under the null model.

$$E[N_{ij}] = \hat{N}_i \frac{\hat{N}_j}{\hat{N}_{..}} \quad (1)$$

where $\hat{N}_{..}$ represents the total number of migrants across the entire network (we analyze separately domestic and international migration). $\hat{N}_i$ is the total number of emigrants leaving the municipality $i$ and $\hat{N}_j$ is the total number of immigrants arriving at location $j$ within the same period.

In the second step, the NC algorithm calculates the difference between the observed edge, the observed migration flow, $\hat{N}_{ij}$ and the expected edge weights $E[N_{ij}]$ to estimate the so-called edge weights' lift. As shown in equation (2), the lift measures how unexpectedly high an edge weight is given the weights of $i$ and $j$.

$$L_{ij} = \frac{\hat{N}_{ij}}{E[N_{ij}]} \quad (2)$$

Since the lift $L_{ij}$ is not symmetric around 1, a transformation to center it around zero, $L'_{ij}$ is applied for easier interpretability.

In the third step, the NC algorithm calculates the variance of the edge weights, which is the variation of the migration flows $V[\hat{N}_{ij}]$. The variance, given by equation (3), is derived from a binomial distribution, as migration flows are treated as a series of independent interactions (i.e., migrating from area $i$ to $j$).

$$V[N_{ij}] = N_{..}P_{ij}(1 - P_{ij}) \tag{3}$$

where $P_{ij}$ is the probability of having a migration flow going from location $i$ to location $j$, relative to all migration flows in the network. $P_{ij}$ is unknown but can be estimated as the observed frequency with which interactions occur, as shown in equation (4).

$$\hat{P}_{ij} = \frac{\hat{N}_{ij}}{\hat{N}_{..}} \tag{4}$$

A common problem in network analysis, especially with sparse networks, is that for certain node pairs $i$ and $j$, the observed migration flow may be zero $\hat{N}_{ij} = 0$. In such cases, its variance will also be zero, $V[N_{ij}] = N_{..}P_{ij}(1 - P_{ij}) = 0$. In other words, when no observed migration occurs between two locations, it becomes difficult to estimate the likelihood of migration flows between them with any precision. The NC algorithm addresses this uncertainty by estimating $P_{ij}$ using a Bayesian framework.

This Bayesian framework uses a Beta distribution to model the probability of migration between two municipalities $P_{ij}$, incorporating prior information about migration patterns. The prior is based on the average number of migrants leaving and arriving at each municipality across the entire migration network, with the parameters $\alpha$ and $\beta$ derived from the mean and variance of those flows. As actual migration data become available, the algorithm updates the prior to form a posterior distribution. This posterior, also modeled by a Beta distribution, refines the initial estimates by incorporating the observed migration data for each municipality $\hat{P}_{ij}$. So the posterior is just a refined version of the initial guess (the prior) after considering all the available data. The posterior distribution also follows a Beta distribution $\hat{P}_{ij}$ ~Beta $[n_{ij} + \alpha, n_{..} - n_{ij} + \beta]$. Where $n_{ij} = 0$ (since there is no observed migration for that node), and the prior parameters $\alpha$ and $\beta$, and $n_{..}$ refer to the same total number of migration flows in the network based on the prior information when dealing with uncertainty.

### *4.1 Our modification of the NC algorithm to calculate changes in the migration network*

Once the posterior distribution for $\hat{P}_{ij}$ is calculated, the original NC algorithm evaluates the expected variance of edge weights to determine which migration flows are statistically significant at time *t*. This is where backboning comes in. Backboning is a strategy used to simplify complex, noisy networks by retaining only the most meaningful connections, in this case, migration flows between two municipalities at time *t*. The algorithm sets a threshold, controlled by the parameter $\delta$, and removes edges that do not meet this level of statistical significance. This helps to highlight the underlying structure of the network by pruning weaker, less reliable connections, leaving only the most relevant flows for further analysis. Our approach deviates from the original NC algorithm in two key ways, as explained next.

Our primary goal is to analyze changes in the migration network over time, comparing two points, *t* and *t+1* (e.g., the 2010 census and the 2015 intercensal survey). Pruning weak edges at a single time point would present a challenge, as it might falsely suggest that certain edges emerge or disappear in the following period, when they may have simply been present but weak. To avoid this, we make a slight adjustment to the NC algorithm by retaining *all* edges in the network, derived from the posterior distribution, instead of filtering based on statistical significance.

Second, we evaluate whether migration flows between two municipalities have increased or decreased over time. This is done by analyzing the edge weight variance to assess the probability that the edge's weight is effectively zero. We derive a posterior distribution for the same edge at different moments in time, *t* and *t+1*, and use the inferred variances to quantify whether the edge weight has significantly changed between these two periods using t-tests. To account for uncertainty, we apply bootstrapping.

We do not use these t-tests to prune statistically significant changes. Instead, the t-tests indicate directional trends, whether migration flows are increasing or decreasing, without removing any edges. A positive t-score reflects an increase in migration flows, while a negative t-score indicates a decrease. We refer to the values of these t-tests as z-scores because they provide a standardized way to measure changes in migration flows, even though they are not bounded between 0 and 1. Like traditional z-scores, they represent deviations from expected values, allowing for comparisons across different networks and time periods.

By retaining all edges, we avoid the problem of artificially emerging edges caused by pruning at one time point *t* but not at another *t+1*, or vice versa. This approach ensures that we can accurately track changes in the migration network, including its weak connections. The standardized z-scores condense changes for each municipality into a single metric, capturing shifts across millions of potential connections with other municipalities over time.

*4.2 Migration network patterns*

Table 1 provides key descriptive statistics at the municipal level, capturing shifts in migration patterns between the 2010 census and the 2015 intercensal survey, as well as between the 2015 survey and the 2020 census. Each of these captures migration that occurred during the five years preceding the respective interview. The table reveals divergent trends. The number of domestic emigrants shows a decline between 2005-2010 and 2010-2015 (-233,999 on average per municipality). In contrast, the z-scores suggest a surge in the domestic emigration network during this period (3.667), indicating an intensification of emigration flows and strengthening of the network. For domestic immigration, both metrics, the change in the number of immigrants (-234.67) and the z-scores (-8.24), show a decline. In contrast, between 2010-2015 and 2015-2020, there was a marked increase in both the number of migrants and their respective z-scores for domestic emigration (205.62 in the number of emigrants and 331.58 in z-scores) and domestic immigration (199.22 and 276.56, respectively), though significant regional variation persists, as illustrated in Fig. 7.

**Table 1**

Descriptives per municipality

| | Mean | Standard Deviation | Observations | Mean | Standard Deviation | Observations |
|---|---|---|---|---|---|---|
| | Change in period 2005-2010 vs 2010-2015 | | | Change in period 2010-2015 vs 2015-2020 | | |
| Change in the number of domestic emigrants leaving municipalities | -233.999 | 3555.338 | 2456 | 205.619 | 2507.785 | 2456 |
| Change in the number of domestic immigrants moving into municipalities | -234.676 | 3120.915 | 2456 | 199.223 | 3451.752 | 2456 |
| Change in the number of Mexicans that returned from the U.S. | -453.003 | 620.070 | 2456 | 99.889 | 304.276 | 2456 |
| Change in the number of Mexicans that returned from countries other than the U.S. | 256.747 | 866.742 | 2456 | -234.098 | 775.761 | 2456 |
| Z-score of change in migration network of domestic emigrants | 3.667 | 8354.301 | 2456 | 331.580 | 7972.509 | 2456 |
| Z-score of change in migration network of domestic immigrants | -8.247 | 5592.579 | 2456 | 276.566 | 7346.668 | 2456 |
| Z-score of change in migration network of return immigration of Mexicans from the U.S. | -155.474 | 367.819 | 2456 | -60.477 | 201.625 | 2456 |
| Z-score of change in migration network of return immigration of Mexicans from countries other than the U.S. | 720.510 | 737.427 | 2456 | -457.170 | 469.265 | 2456 |
| | Change in years 2005 vs 2010 | | | Change in years 2010 vs 2015 | | |
| Change in the number of people who left Mexico for the U.S. | -24.789 | 107.296 | 2456 | 4.234 | 75.349 | 2433 |
| Change in the number of people who left Mexico for countries other than the U.S. | -0.977 | 24.365 | 2433 | 3.447 | 27.363 | 2433 |
| Z-score of change in migration network of emigration to the U.S. | -98.030 | 181.428 | 2456 | 91.466 | 170.718 | 2433 |
| Z-score of change in migration network of emigration to non-U.S. countries | -10.726 | 70.918 | 2433 | 42.398 | 90.263 | 2433 |
| Change in the number of homicides | 6.612 | 81.470 | 2394 | -2.099 | 74.575 | 2394 |
| Change in the homicide rate | 12.108 | 77.131 | 2393 | -5.049 | 72.859 | 2394 |
| Change in the gini index | 0.017 | 0.050 | 2453 | -0.041 | 0.043 | 2445 |
| Change in the poverty rate | -0.153 | 0.100 | 2442 | -0.016 | 0.055 | 2456 |
| Change in night-time light per capita | 0.003 | 0.022 | 2455 | -0.044 | 0.048 | 2456 |
| Change in the average price of diesel and gasoline divided by the municipality's average distance to the nearest oil pipeline | 0.149 | 2.637 | 2456 | 0.552 | 9.631 | 2456 |
| Change in the price of gasoline in real pesos | 0.951 | 0.012 | 2456 | 2.364 | 0.013 | 2456 |

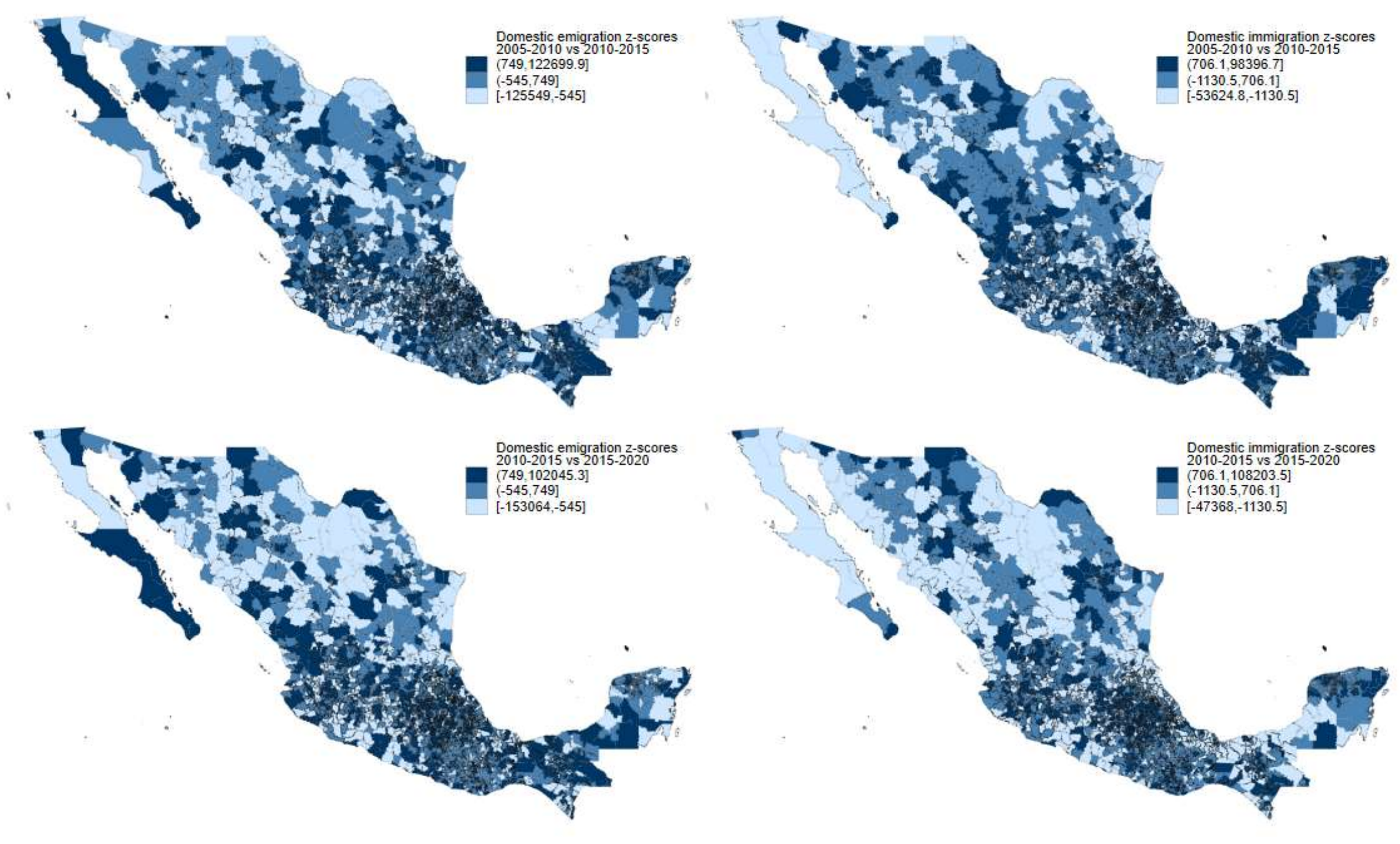

**Fig. 7.** Domestic migration z-scores obtained with the NC method

About 90% of Mexico's international migration is directed towards the U.S. Table 1 also provides the average changes in these migration flows per municipality, rather than national totals. Between 2005–2010 and 2010–2015, migration to the U.S. declined sharply, as evidenced by a fall in the number of emigrants (-24.79) and z-scores (-98.03). However, from 2010–2015 to 2015–2020, there was a noticeable rebound in emigration, reflected in the increase in both the number of emigrants (4.23) and z-scores (91.47). Return migration from the U.S. of Mexican nationals also saw a significant drop between 2005–2010 and 2010–2015 (-453.00 and -155.47, respectively). While the period from 2010–2015 to 2015–2020 showed a deceleration in this decline, with a smaller reduction in the number of immigrants (99.89) and z-scores (-60.48), substantial regional heterogeneity persists (Fig. 8).

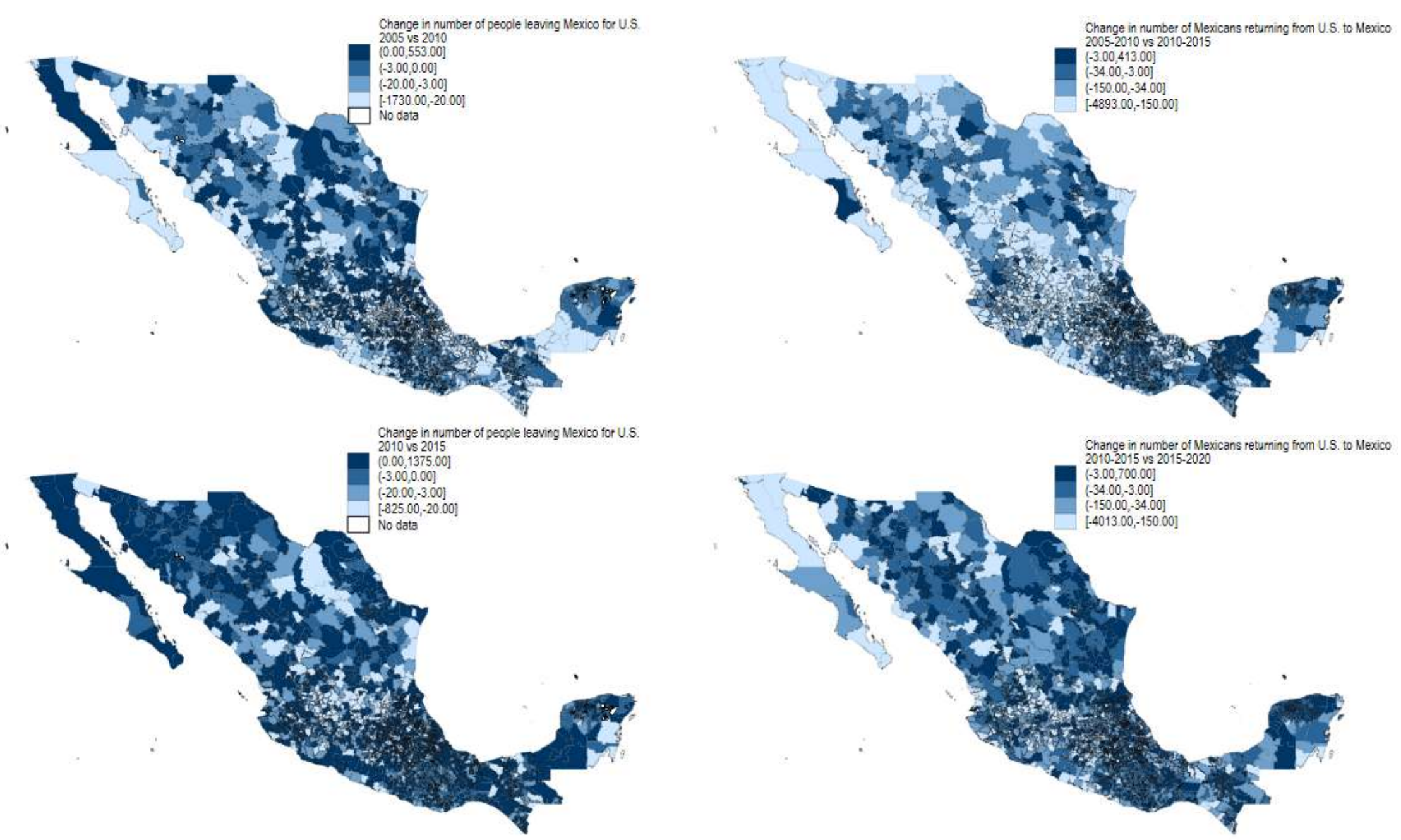

**Fig. 8.** International migration z-scores obtained with the NC method

The changes in Mexico's domestic migration network can also be visualized. For instance, Fig. 9 compares migration flows between the 2015 intercensal survey and the 2020 census. The nodes represent municipalities, and the edges reflect migration connections, estimated using the NC backboning algorithm. Municipalities are color-coded: green indicates increased immigration connectivity, yellow reflects neutral changes, orange shows moderate decline, and red marks a strong decline. Notably, the network is not spatially constrained, significant migration flows occur between distant regions, suggesting that proximity is not the primary driver of immigration.

The network depicts several municipalities in green, indicating that immigration is not concentrated in a few major urban centers but is spread across a wide range of areas (Fig. 9). To improve readability and avoid clutter, we omit the names of most municipalities from the figure, highlighting only a few notable cases. For instance, municipalities like Tijuana and Juárez (in Chihuahua) are expanding their domestic immigration network by receiving flows from wider areas. Other areas near Mexico City such as Tlalnepantla and Ecatepec (both in Estado de México), have reduced their immigration network. In the following section, we examine the potential impact of rising homicide rates on the observed changes in the emigration and immigration networks.

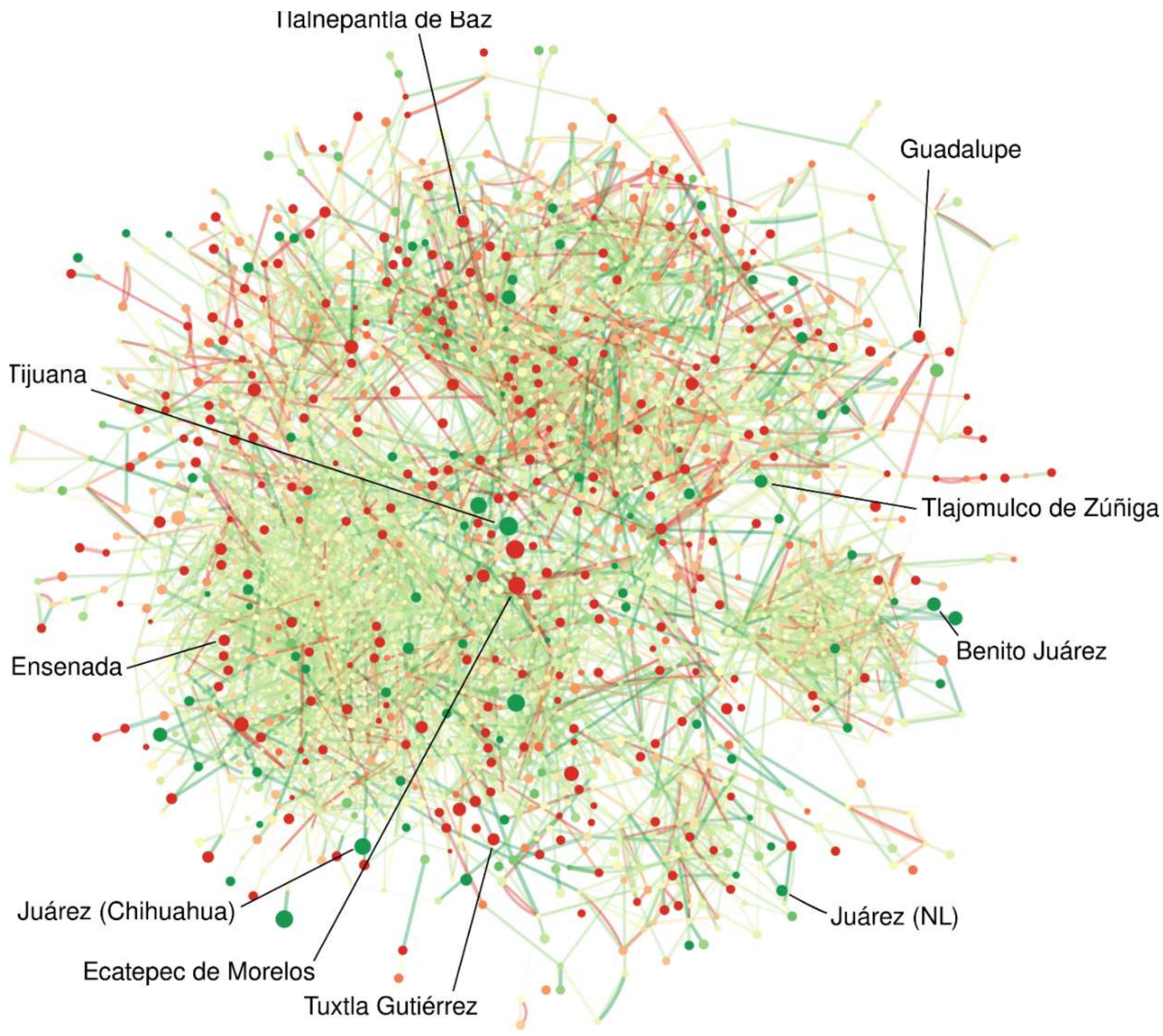

**Fig. 9.** Change in the domestic immigration comparing migration flows between the periods 2010–2015 and 2015–2020

## 5. Empirical strategy and findings

Once we have quantified the changes in migration networks, we proceed to identify to what extent violence drove those changes. We begin by estimating a panel fixed effects regression, as shown in equation (5).

$$\Delta m_{it} = \beta_1 \Delta Homicides_{it} + \beta_2 \Delta X_{it} + \beta_3 T_t + \mu_i + \varepsilon_{it} \quad (5)$$

where the dependent variable $\Delta m_{it}$ denotes the standardized z-scores, which measure changes in the migration network for municipality $i$ during time $t$ (i.e., change in z-scores when comparing census data during 2005-2010 vs. 2010-2015 and 2010-2015 vs. 2015-2020). We use four dependent variables separately. These are the z-scores for domestic emigration, domestic immigration, international emigration to the U.S., and return migration of Mexicans from the U.S. Separately, we also analyze the change in the number of migrants, whether emigrants or immigrants, domestic or international.

$\Delta Homicides_{it}$ denotes the change in the homicide rate during the quinquennial periods compared (i.e., 2005 vs. 2010 and 2010 vs. 2015). The vector $\Delta X_{it}$ denotes the change in the poverty rate, and the Gini coefficient during the periods compared (i.e., 2005 vs. 2010 and 2010 vs. 2015). We also add the changes in annual nighttime light-per-capita, over the periods 2001-2005, 2006-2010, and 2011-2015, as a proxy for wealth. The specification includes a dummy variable comparing the quinquennial terms being compared $T_t$ (equal to 1 for the first quinquennial comparison 2005-2010 vs. 2010-2015 and zero otherwise), as well as municipality fixed effects $\mu_i$ . The regression residual is denoted by $\varepsilon_{it}$. The robust standard errors are clustered at the municipality level.

*5.1 Instrumental variables and validity of instruments*

While the panel fixed effects regression controls for time-invariant differences across municipalities, it is highly unlikely to fully address endogeneity arising from time-varying factors that simultaneously affect homicide rates and migration. In particular, the location of battles between drug trafficking criminal organizations driving much of the homicides is not random. These criminal groups have violently contested control over drug smuggling routes and areas where large-scale fuel theft is feasible. These areas may also attract or deter domestic and international migrants for reasons unrelated to violence, such as changes in institutional conditions or economic opportunities. As a result, changes in homicides and migration networks could be jointly driven by unobserved local shocks, biasing the fixed effects estimates.

If unobserved economic shocks drive both homicides and migration, the estimates may be upward biased, overstating the likely deterrent effect of violence on migration. Conversely, if violence depresses local economic activity and discourages migration simultaneously, the estimates may be downward biased, understating the true effect of the rise in homicides. To address this concern, we implement an instrumental variables (IV) strategy, combining it with municipality and year fixed effects to isolate plausibly exogenous variation in violence.

Our identification relies on two instruments: (i) the national changes in fuel prices, which have been driven by exogenous national policy shifts and the gradual elimination of fuel subsidies following the financial collapse of the national oil company, PEMEX, and (ii) the differences among municipalities in their exposure to oil theft, measured by the interaction between these price changes and the inverse distance from each municipality's centroid to the nearest oil pipeline.

These instruments capture exogenous variation in the likelihood of territorial conflicts among criminal organizations, thereby affecting changes in local homicide rates. Large-scale oil theft occurs directly from these pipelines carrying gasoline and diesel. That is why areas located close to this infrastructure have been heavily contested by criminal groups due to the high value of fuel theft revenues.

The exclusion restriction requires that the distance to the nearest pipeline and the change in fuel affect homicides but do not directly influence changes in migration networks or flows. This assumption is plausible for at least three key reasons. First, pipeline infrastructure is buried underground, was built primarily during the 1950s–60s, and serves only to transport oil from national extraction points across the country. So, proximity to a pipeline does not provide any economic advantages or disadvantages that would systematically influence migration decisions. Second, retail fuel prices have been centrally set by the federal government long before the war on drugs. These changes were linked to international market conditions and national subsidies, which have remained largely uniform nationwide. Only very minor zone-level adjustments have been allowed in a few remote or high-cost-to-reach areas affecting primarily only two states.[4] However, these differences are in so few states, and so small (see Table 1) that they do not generate any meaningful variation in local economic conditions relevant to migration. Third, while some stolen fuel is sold locally, the vast majority is funneled through legal fuel stations and sold at federally regulated prices (Jones and Sullivan 2019). In practice, profits from oil theft largely accrue to criminal organizations rather than being redistributed to local populations (Battiston et al. 2024; Gutiérrez-Romero and Iturbe 2024; Jones and Sullivan 2019).

To implement our instrumental variables strategy, we first estimate the relationship between the instruments and changes in homicide rates, relevant for our period of analysis. This first-stage specification is presented in equation (6):

$$\Delta Homicides_{it} = \mu_{1i} Z_{it} + \mu_{2i} \Delta X_{it} + \mu_{3i} \Delta T_t + \mu_i + u_{it} \quad (6)$$

[4] Since 2017, more regional variance has been introduced due to the full market deregulation of fuel and the removal of fuel subsidies; however the changes in fuel prices do not cover this more recent period.

As mentioned earlier, our likely endogenous variable, the change in homicide rates, is measured as the difference between 2005 and 2010, and between 2010 and 2015. This timing is to explain likely changes in our main outcome: the change in migration networks, derived from comparing censuses. The vector $Z$ includes two instruments. The first is the five-year change in the price of premium gasoline, the fuel most closely tied to international market prices, measured as the difference between 2005 and 2010, and between 2010 and 2015. The second is the interaction between the change in the average price of fuel (including premium, Magna, and diesel) and the inverse distance from each municipality's centroid to the nearest oil pipeline. For this instrument, fuel price changes are measured between 2004 and 2009, and between 2009 and 2014.

We use slightly different periods for capturing changes in fuel prices for two reasons: (i) to reflect that fuel price increases may lead to a rise in homicide rates with a lag, particularly in municipalities near pipelines, where rising prices increase the value of fuel theft; and (ii) to account for the growing perception during this period that fuel price shocks (known in Mexico as *gasolinazos*), although unannounced and unexpected, were becoming more frequent, thereby increasing the incentives for large-scale oil theft.

All fuel prices are adjusted to real terms using the national consumer price index from INEGI. Gasoline and diesel prices are sourced from the Comisión Federal de Competencia Económica (COFECE). The distances to oil pipelines are our estimates, based on geospatial data from PEMEX.

To test whether the first-stage is sufficiently strong, we need to examine whether municipalities located closer to pipelines experienced larger increases in homicide rates than those farther away, particularly during periods of rising fuel prices. As shown in the next subsection, our results suggest that the instruments are reasonably strong. The standard endogeneity tests reveal that the simple panel fixed specifications are downward biased as they do not capture the true impact of the rise in homicide rates, and we should therefore prefer the IV specification.

The second-stage IV panel fixed effects specification, denoted by equation (7), estimates the impact of the instrumented change in the homicide rate on the z-score (which denotes the change in the migration network).

$$\Delta m_{it} = \kappa_1 \Delta \widehat{Homicides}_{it} + \kappa_2 \Delta X_{it} + \kappa_3 T_t + \kappa_i + e_{it} \qquad (7)$$

where $\kappa_1$ is the regression coefficient of the instrumented change in the homicide rate. The residuals are denoted by $\kappa_i$ and $e_{it}$.

*5.2 Impact of violence on the migration network*

We begin by examining the impact of homicide rates on domestic and international migration networks. Table 2, columns 1–4, reports the panel fixed effects estimates. The results indicate that rising homicide rates are associated with declines in both the network of domestic migration and emigration to the U. S., as measured by the respective z-scores. However, none of these associations are statistically significant. As discussed earlier, these estimates may be biased due to potential endogeneity.

**Table 2**

Change in homicide rate and z-standardized migration scores

| | (1) | (2) | (3) | (4) | (5) | (6) | (7) | (8) |
|---|---|---|---|---|---|---|---|---|
| | Panel fixed effects | | | | Second-stage IV panel fixed effects | | | |
| | Change in domestic emigration z-scores | Change in domestic immigration z-scores | Change in international emigration (leaving for U.S.) z-scores | Change in international immigration (Mexicans return from U.S.) z-scores | Change in domestic emigration z-scores | Change in domestic immigration z-scores | Change in international emigration (leaving for U.S.) z-scores | Change in international immigration (Mexicans return from U.S.) z-scores |
| Change in the homicide rate | -0.146 | -0.398 | -0.065 | -0.121 | 83.47** | 49.91** | 2.032** | 3.229 |
| | (0.718) | (0.770) | (0.054) | (0.112) | (34.41) | (21.79) | (0.909) | (2.242) |
| Other controls | Yes | Yes | Yes | Yes | Yes | Yes | Yes | Yes |
| Municipalities fixed effects | Yes | Yes | Yes | Yes | Yes | Yes | Yes | Yes |
| Time fixed effects | Yes | Yes | Yes | Yes | Yes | Yes | Yes | Yes |
| Observations | 4,736 | 4,736 | 4,700 | 4,736 | 4,736 | 4,736 | 4,700 | 4,736 |
| Number of municipalities | 2,368 | 2,368 | 2,350 | 2,368 | 2,368 | 2,368 | 2,350 | 2,368 |
| F test of excluded instruments: | | | | | 15.16 | 15.16 | 15.53 | 15.16 |
| Prob > F | | | | | 0.00 | 0.00 | 0.00 | 0.00 |
| Stock-Yogo weak ID test critical values: 15% maximal IV size | | | | | 11.59 | 11.59 | 11.59 | 11.59 |
| Hansen J statistic (overidentification test of all instruments) | | | | | 2.60 | 2.26 | 1.77 | 1.16 |
| Chi-sq(1) P-val | | | | | 0.11 | 0.13 | 0.18 | 0.28 |
| Endogeneity test of endogenous regressors: | | | | | 5.82 | 7.43 | 6.30 | 2.41 |
| Chi-sq(1) P-val | | | | | 0.02 | 0.01 | 0.01 | 0.12 |

Other controls include changes in the poverty rate, annual nighttime light-per-capita, and the Gini coefficient. Robust standard errors, clustered at the municipality level, are in parentheses. Significance levels *** p<0.01, ** p<0.05, * p<0.1.

To address concerns about endogeneity, we turn to an instrumental variables (IV) panel fixed effects strategy. Table 3 presents the first-stage results corresponding to the second-stage IV regressions shown in Table 2, columns 5–8. There is a minor difference in sample size for the international emigration regression (column 3), resulting in slight variation in the first-stage estimates. However, the instruments consistently explain variation in homicide rates. Both instruments (the five-year change in premium gasoline prices and the interaction between average fuel prices and inverse distance to the nearest oil pipeline) are strong predictors of changes in homicide rates and statistically significant across specifications.

The interaction term captures the idea that municipalities closer to pipelines are more exposed to fuel theft-driven violence, particularly when fuel prices rise. The higher this interaction (i.e., higher oil prices and closer proximity to pipelines), the more homicide rates rise. The other instrument, the five-year change in gasoline prices, also reveals that rises in gasoline prices are linked to higher homicide rates.

**Table 3**

First-stage regression of Table 2

| | (1) | (2) | (3) | (4) |
|---|---|---|---|---|
| Second-stage IV regression's dependent variable: | Change in domestic emigration z-scores | Change in domestic immigration z-scores | Change in international emigration (leaving for U.S.) z-scores | Change in international immigration (Mexicans return from U.S.) z-scores |
| First-stage regression's dependent variable: | Change in the homicide rate | Change in the homicide rate | Change in the homicide rate | Change in the homicide rate |
| Change in the average price of diesel and gasoline divided by the municipality's average distance to the nearest oil pipeline | 0.103** | 0.103** | 0.108** | 0.103** |
| | (2.27) | (2.27) | (2.42) | (2.27) |
| Change in the price of gasoline in real pesos | 370.5*** | 370.5*** | 372.7*** | 370.5*** |
| | (5.33) | (5.33) | (5.36) | (5.33) |
| Change in the Gini index | -37.22 | -37.22 | -50.07 | -37.22 |
| | (-0.57) | (-0.57) | (-0.77) | (-0.57) |
| Change in the poverty rate | -42.52 | -42.52 | -51.93 | -42.52 |
| | (-1.30) | (-1.30) | (-1.63) | (-1.30) |
| Change in the nighttime light-per-capita | 605.4*** | 605.4*** | 590.4** | 605.4*** |
| | (3.35) | (3.35) | (3.24) | (3.35) |
| Municipality fixed effects | Yes | Yes | Yes | Yes |
| Year fixed effects | Yes | Yes | Yes | Yes |
| Number of municipalities | 4736 | 4736 | 4700 | 4736 |

Robust standard errors, clustered at the municipality level, in parentheses.

Significance levels *** p<0.01, ** p<0.05, * p<0.1.

The first-stage F-statistics exceed the conventional threshold of 15, indicating strong instrument relevance (Table 2, columns 5–8). The Cragg–Donald Wald F-statistics and Hansen J overidentification tests, reported at the bottom of Table 2, suggest that the instruments are both relevant and valid. Endogeneity tests, also shown in Table 2, reject the null of exogeneity for domestic emigration, domestic immigration, and international emigration to the U.S., indicating that IV estimates should be preferred. At any rate, both the fixed effects and IV estimates suggest that increases in homicide rates did not significantly affect the return migration network.

For return migration from the U.S., the endogeneity test yields a p-value of 0.12, providing no evidence of significant endogeneity in this case. The instruments remain relevant, as they explain substantial variation in homicide rates, drawing on fuel price changes that reflect national and international market dynamics, including those in the U.S. driving international fuel prices. While unobserved shocks may bias estimates in other migration outcomes by simultaneously affecting violence and migration decisions, the data do not indicate a strong correlation between such unobserved shocks and return migration. As a result, although the IV strategy remains valid and supported by relevant instruments, the panel fixed effects estimator is likely sufficient for identifying the effect of homicides on return migration in this specification.

The second-stage IV panel fixed effects estimation, Table 2, shows how rising homicide rates have reshaped migration networks, particularly within Mexico. A one-unit increase in homicide rates significantly increased both domestic emigration (83.47) and domestic immigration (49.91) networks, proxied by changes in their z-scores. There is a key distinction, though: the domestic emigration network is becoming stronger than the immigration network, meaning that more people are leaving than arriving, and from a wider variety of areas. Importantly, the impact of rising homicide rates on the domestic emigration network is much stronger than on the international emigration network to the U.S., where the effect is positive but smaller (2.03), as seen in Table 2, column 7. These findings suggest that, while higher homicide rates are driving some people to leave Mexico altogether, reflected in a strengthened emigration network to the U.S., the overwhelming impact is felt within the country. The weak and statistically insignificant effect on the return migration network from the U.S. (3.23) suggests that rising homicide rates are not impacting the strength of this network (Table 2, column 8).

*5.3 Impact of violence on the net change of the number of migrants*

All previous studies on the impact of violence on migration in Mexico have focused on changes in net flows, tracking the number of people moving into or out of areas. While informative, these net flows overlook the complexity of how violence reshapes the intricate domestic and international migration networks across all municipalities. This is why we chose the NC algorithm, which allows for a more detailed analysis of these bipartite networks. But what if, as in previous research, we had analyzed only net migrant flows? Would we have arrived at the same mixed results as earlier research?

To our knowledge, previous studies have not fully examined the impact of violence on the return of Mexicans from the U.S.[5] Earlier studies have not evaluated the impact of homicides on migration over such an extended period either, including through 2020. Thus, our findings offer insights into whether the effects of rising homicide rates on migration are consistent when focusing on changes in migration flows, while also revealing how violence has affected migration networks.

---

[5] Bucheli, Fontenla, and Wadell (2019) analyze return migration from the U.S. They use instrumental variables to identify the effect of return migration on homicide rates in Mexico during the period 2011-2013, a different issue from the research question analyzed here.

To evaluate the impact on migrant flows, we re-estimate our analysis using the change in the number of people migrating into and out of each municipality (both domestically and to the U.S.) as the dependent variables. Each census records flows from the preceding five years, resulting in two periods of analysis: 2005–2010 and 2010–2015 for the first census comparison, and 2010–2015 and 2015–2020 for the second.

The main results are presented in Table 4. The panel fixed effects results, columns 1-4, suggest that both emigration and immigration are negatively associated with homicide rates, but these results are not statistically significant. For international migration, we see a statistically significant, albeit small, negative effect on emigration to the U.S. The effect of homicide rates on return migration from the U.S. is also negative, but not significant.

Given the consensus in the literature about endogeneity concerning homicide rates and migration, we also present in Table 4, columns 5-8, the second-stage IV panel fixed effects regressions. To ensure methodological consistency, we use exactly the same set of instruments as in our network analysis.

**Table 4**

Change in the homicide rate and change in the number of migrants

| | (1) | (2) | (3) | (4) | (5) | (6) | (7) | (8) |
|---|---|---|---|---|---|---|---|---|
| | Panel fixed effects | | | | Second-stage IV panel fixed effects | | | |
| | Change in number of domestic emigration | Change in number of domestic immigration | Change in number of international emigration (leaving for U.S.) | Change in number of international immigration (Mexicans return from U.S.) | Change in number of domestic emigration | Change in number of domestic immigration | Change in number of international emigration (leaving for U.S.) | Change in number of international immigration (Mexicans return from U.S.) |
| Change in the homicide rate | -0.229 | -0.988 | -0.054* | -0.057 | 39.58** | 33.33** | 0.793* | -1.779** |
| | (0.387) | (0.874) | (0.032) | (0.063) | (17.03) | (14.75) | (0.461) | (0.877) |
| Other controls | Yes | Yes | Yes | Yes | Yes | Yes | Yes | Yes |
| Municipalities fixed effects | Yes | Yes | Yes | Yes | Yes | Yes | Yes | Yes |
| Time fixed effects | Yes | Yes | Yes | Yes | Yes | Yes | Yes | Yes |
| Observations | 4,774 | 4,774 | 4,738 | 4,774 | 4,736 | 4,736 | 4,700 | 4,736 |
| Number of municipalities | 2,394 | 2,394 | 2,376 | 2,394 | 2,368 | 2,368 | 2,350 | 2,368 |
| F test of excluded instruments: | | | | | 15.16 | 15.16 | 15.53 | 15.16 |
| Prob > F | | | | | 0.00 | 0.00 | 0.00 | 0.00 |
| Stock-Yogo weak ID test critical values: 15% maximal IV size | | | | | 11.59 | 11.59 | 11.59 | 11.59 |
| Hansen J statistic (overidentification test of all instruments) | | | | | 2.12 | 0.54 | 1.95 | 0.49 |
| Chi-sq(1) P-val | | | | | 0.15 | 0.46 | 0.16 | 0.49 |
| Endogeneity test of endogenous regressors: | | | | | 5.45 | 10.25 | 4.24 | 4.20 |
| Chi-sq(1) P-val | | | | | 0.02 | 0.00 | 0.04 | 0.04 |

Other controls include changes in the poverty rate, annual nighttime light-per-capita, and the Gini coefficient. Robust standard errors, clustered at the municipality level, are in parentheses. Significance levels *** p<0.01, ** p<0.05, * p<0.1.

Table 5 reports the first-stage regressions for the IV specifications used to estimate the effects of homicide rates on changes in the number of migrants. These first-stage results are identical to those shown in Table 3, which corresponds to the specifications using migration network z-scores as outcomes. The endogenous regressor (change in homicide rates), instruments, controls, fixed effects, and sample size are all the same across specifications. We present the first-stage estimates again to make clear that differences in second-stage results are not driven by changes in sample composition or model specification. In all columns, the instruments (the five-year change in gasoline prices and the five-year change in the average price of oil relative to pipeline proximity) remain statistically significant and have the expected signs.

**Table 5**

First-stage regression for Table 4

| | (1) | (2) | (3) | (4) |
|---|---|---|---|---|
| Second-stage IV regression's dependent variable: | Change in number of domestic emigration | Change in number of domestic immigration | Change in number of international emigration (leaving for U.S.) | Change in number of international immigration (Mexicans return from U.S.) |
| First-stage regression's dependent variable: | Change in the homicide rate | Change in the homicide rate | Change in the homicide rate | Change in the homicide rate |
| Change in the average price of diesel and gasoline divided by the municipality's average distance to the nearest oil pipeline | 0.103** | 0.103** | 0.108** | 0.103** |
| | (2.27) | (2.27) | (2.42) | (2.27) |
| Change in the price of gasoline in real pesos | 370.5*** | 370.5*** | 372.7*** | 370.5*** |
| | (5.33) | (5.33) | (5.36) | (5.33) |
| Change in the gini index | -37.22 | -37.22 | -50.07 | -37.22 |
| | (-0.57) | (-0.57) | (-0.77) | (-0.57) |
| Change in the poverty rate | -42.52 | -42.52 | -51.93 | -42.52 |
| | (-1.30) | (-1.30) | (-1.63) | (-1.30) |
| Change in the night-time light per capita | 605.4*** | 605.4*** | 590.4** | 605.4*** |
| | (3.35) | (3.35) | (3.24) | (3.35) |
| Municipality fixed effects | Yes | Yes | Yes | Yes |
| Year fixed effects | Yes | Yes | Yes | Yes |
| Number of municipalities | 4736 | 4736 | 4700 | 4736 |

Robust standard errors, clustered at the municipality level, in parentheses.

Significance levels *** p<0.01, ** p<0.05, * p<0.1.

Table 4, columns 5–8, presents the second-stage IV estimates alongside the corresponding panel fixed effects results. The table also reports key first-stage diagnostics: the F-statistic for excluded instruments, the Cragg–Donald Wald F-statistic, and the Hansen J test of overidentifying restrictions. These tests confirm that the instruments are relevant and valid, strongly correlated with the endogenous regressor and plausibly exogenous. The endogeneity tests reported in Table 4 reject the null of exogeneity for domestic and international migration, including return migration from the U.S., reinforcing the need to use the IV approach.

The second-stage IV panel fixed effects estimator reveals that homicide rates significantly impact migration patterns at the municipality level. For domestic migration, a one-unit increase in homicide rates leads to a statistically significant increase in the number of domestic emigrants (39.58) and domestic immigrants (33.33). This suggests that rising violence does not deter people from arriving, potentially for economic opportunities that these municipalities may offer, albeit in net terms more people are leaving. These IV findings support our z-score analyses. They suggest that during the period analyzed from 2005 to 2020, overall the number of domestic emigrants and immigrants has increased, but emigration increased further.

Now we turn to international migration to the U.S. For return migration, the effect is smaller: a one-unit increase in homicide rates corresponds to a statistically significant decrease in the number of Mexicans returning from the U.S. (-1.779). For emigration, a one-unit increase in the five-year change in homicide rates results in a modest but statistically significant rise in the number of non-transit emigrants to the U.S. (0.793). Comparing these trends with the network analysis reveals that, while the structure of the return migration network has remained stable, fewer people are returning because of violence. Meanwhile, the rise in both emigration flows and network strength due to increasing homicide rates suggests that violence acts as a push factor driving migration to the U.S.

The direction of the impact of homicide rates on the number of people emigrating abroad aligns with the findings of Orozco-Aleman and Gonzalez-Lozano (2018) and Daniele, Le Moglie, and Masera (2023). However, these results contrast with those of Basu and Pearlman (2017), who found that violence reduced the number of people migrating to the U.S., likely due to the rising costs of emigration. The differences in the impact on U.S.-bound emigration may be attributed to variations in data, methodologies, instruments, and periods analyzed, highlighting how these factors can shape the interpretation of violence's influence on migration flows.

Despite differences with respect to other studies, as previously stated, the changes in z-scores offer a deeper understanding of underlying shifts in the migration network compared to simply tracking the number of migrants. Our network analysis captures the dynamics across millions of potential origin-destination paths, both within Mexico and abroad.

Another important policy question is ultimately how many migrants nationwide were displaced by violence, and what proportion of total migration this represents. We address this by calculating the total change in the migrant population between 2005 and 2020, captured by the censuses and the intercensal survey analyzed. Although these sources do not account for within-municipality migration or multiple moves by the same individual, they still offer valuable insights into broader migration trends.

We obtain total migration flows nationwide by using the IV results from columns 5 and 8 in Table 4. We multiply the estimated coefficients by the average change in homicide rates between 2005 and 2020 (11.51) and the number of municipalities in the country (2,454). This calculation suggests that over the period analyzed (2005–2020), the rise in homicide rates contributed to an increase of approximately 1,117,968 domestic emigrants and 941,434 domestic immigrants. Additionally, it led to 22,398 more emigrants to the U.S. and a reduction of 50,249 Mexicans returning from the U.S. to more violent Mexican municipalities.

That means that the rise in homicide rates between 2005 and 2020 contributed about 6% of total domestic emigration, 5% of domestic immigration, 1% of international emigration to the U.S., and a decline in return migration from the U.S. of 3%.

*5.4 Impact of violence on migration to non-U.S. countries*

A further contribution of our analysis is to estimate the impact of rising homicide rates in Mexico on both the migration network and the number of emigrants to countries other than the U.S. Since international migration to non-U.S. destinations and rising homicide rates in Mexico may be endogenous, we apply the same instrumental variables (IV) strategy used in earlier sections to isolate exogenous variation in local violence. Endogeneity may arise if unobserved shocks (such as economic conditions that increase incentives to migrate abroad) also lead to higher local violence, for example, by increasing the returns to illicit activity. To address this, we use the same two instruments: (1) the five-year change in premium gasoline prices, and (2) the five-year change in the average price of fuel (premium, Magna, and diesel) divided by the distance from each municipality's centroid to the nearest oil pipeline. These instruments capture plausibly exogenous variation in local violence driven by changes in the profitability of large-scale fuel theft, especially in municipalities near pipeline infrastructure. Although higher fuel prices could, in principle, affect the cost of living and migration incentives, Mexican fuel prices are tied to international markets. As such, fuel price shocks also raise prices in destination countries and therefore should not meaningfully alter the relative economic incentives to migrate abroad. This supports the plausibility of the exclusion restriction.

Table A.1 in the Appendix presents the first-stage IV results for the main outcomes of non-U.S. network migration in columns 1 and 2, along with the flow of emigrants and immigrants from non-U.S. countries in columns 3 and 4. These results closely align with those shown earlier in Tables 3 and 5, as the endogenous variable (the increase in local homicide rates), instruments, and controls remain unchanged. There are minor differences in sample size due to a few municipalities with no observed migration; however, both instruments continue to be statistically significant, strong, and have the expected signs. The instruments are strong, with the F-statistics for the excluded instruments exceeding conventional thresholds. Additionally, the Hansen J test indicates no evidence against the validity of the instruments (see Tables 6 and 7).

We find no evidence of endogeneity in the emigration network to non-U.S. countries or in the total number of emigrants heading to those destinations (see Tables 6 and 7, columns 3). This finding suggests that the fixed effects estimates are more reliable. These estimates indicate that rising homicide rates had no statistically significant impact on the international emigration network to non-U.S. destinations or on the number of emigrants departing for those countries (Tables 6 and 7, column 1). This outcome may not be surprising, given that the majority of international emigration from Mexico is directed toward the U.S. Additionally, our earlier results (Tables 2 and 4) demonstrate that while violence slightly increased international migration and strengthened its network, this effect was limited to migration toward the U.S.

**Table 6**

Change in the homicide rate and international migration network to non-U.S. countries

| | (1) | (2) | (3) | (4) |
|---|---|---|---|---|
| | Panel fixed effects | | Second-stage IV panel fixed effects | |
| | Change in z-scores for emigration to non-U.S. countries | Change in z-scores for returning Mexican immigrants from non-U.S. countries | Change in z-scores for emigration to non-U.S. countries | Change in z-scores for returning Mexican immigrants from non-U.S. countries |
| Change in the homicide rate | -0.0365 | 0.467 | 0.212 | -7.574** |
| | (0.0223) | (0.437) | (0.207) | (3.138) |
| Other controls | Yes | Yes | Yes | Yes |
| Municipalities fixed effects | Yes | Yes | Yes | Yes |
| Time fixed effects | Yes | Yes | Yes | Yes |
| Observations | 4,726 | 4,700 | 4,762 | 4,736 |
| Number of municipalities | 2,376 | 2,350 | 2,394 | 2,368 |
| F test of excluded instruments: | | | 15.53 | 15.16 |
| Prob > F | | | 0.00 | 0.00 |
| Stock-Yogo weak ID test critical values: 15% maximal IV size | | | 11.59 | 11.59 |
| Hansen J statistic (overidentification test of all instruments) | | | 3.43 | 0.37 |
| Chi-sq(1) P-val | | | 0.06 | 0.54 |
| Endogeneity test of endogenous regressors: | | | 1.10 | 9.36 |
| Chi-sq(1) P-val | | | 0.29 | 0.00 |

Other controls include changes in the poverty rate, annual nighttime light-per-capita, and the Gini coefficient. Robust standard errors, clustered at the municipality level, are in parentheses. Significance levels *** $p<0.01$, ** $p<0.05$, * $p<0.1$.

We find evidence of endogeneity in both the return migration network of Mexicans from non-U.S. countries and the total number of Mexican returnees (see Tables 6 and 7, column 4), which indicates that the IV estimates should be preferred. Table 6 shows that rising homicide rates resulted in a statistically significant contraction of the return migration network, with a coefficient of -7.57. In contrast, Table 7 reveals no statistically significant change in the total number of returnees. This suggests that, while the same number of return migrants is being received, the strength of the network has weakened, resulting in a narrower set of source countries for these returnees.

**Table 7**

Change in homicide rate and number of international migrations to non-U.S. countries

| | (1) | (2) | (3) | (4) |
|---|---|---|---|---|
| | Panel fixed effects | | Second-stage IV panel fixed effects | |
| | Change in number of emigrants to non-U.S. countries | Change in number of returning Mexican immigrants from non-U.S. countries | Change in number of emigrants to non-U.S. countries | Change in number of returning Mexican immigrants from non-U.S. countries |
| Change in the homicide rate | -0.00874 | 0.119 | 0.475* | -7.250 |
| | (0.00579) | (0.144) | (0.266) | (4.649) |
| Other controls | Yes | Yes | Yes | Yes |
| Municipalities fixed effects | Yes | Yes | Yes | Yes |
| Time fixed effects | Yes | Yes | Yes | Yes |
| Observations | 4,726 | 4,700 | 4,762 | 4,736 |
| Number of municipalities | 2,376 | 2,350 | 2,394 | 2,368 |
| F test of excluded instruments: | | | 15.53 | 15.16 |
| Prob > F | | | 0.00 | 0.00 |
| Stock-Yogo weak ID test critical values: 15% maximal IV size | | | 11.59 | 11.59 |
| Hansen J statistic (overidentification test of all instruments) | | | 1.37 | 0.87 |
| Chi-sq(1) P-val | | | 0.24 | 0.35 |
| Endogeneity test of endogenous regressors: | | | 0.54 | 4.68 |
| Chi-sq(1) P-val | | | 0.46 | 0.03 |

Other controls include changes in the poverty rate, annual nighttime light-per-capita, and the Gini coefficient. Robust standard errors, clustered at the municipality level, are in parentheses. Significance levels *** p<0.01, ** p<0.05, * p<0.1.

## 6. Robustness check: highway network

As a robustness check, we analyze how violence in Mexico reshapes population dynamics by tracking changes in inter-municipal highway traffic. If rising homicide rates drive residents away, we should observe a lasting drop in vehicle flows linking affected municipalities to the broader highway network. By concentrating exclusively on long-term shifts in inter-municipal traffic, rather than on short-term reductions in local trips, which residents may temporarily curtail out of fear of crime, we offer independent evidence that may lend credence to our migration network findings.

For our analysis, we compare annualized average daily counts of cars, buses, and freight trucks traversing each municipality in 2005 and 2015, capturing traffic patterns before and after the escalation of Mexico's war on drugs. To avoid distortions from COVID-19 mobility constraints and contagion fears, we cap our analysis at 2015 rather than extend it through 2020. Because the 2020 census asks respondents to report moves since 2015, ending our traffic comparison in that year aligns our highway-traffic measure with the census's migration reference period without overlapping pandemic-affected years. Building this dataset took months of work, georeferencing every segment of the national highway network and extracting traffic-count figures from PDF reports published by the Ministry of Communications and Transport.

We quantify how each municipality's connectivity to the rest of the country changed between 2005 and 2015, using annualized daily vehicle trips in all directions (inbound, outbound, or both). To this end, we apply the modified NC algorithm, which retains all links regardless of strength, and computes a z-score for each municipality, indicating whether its network connectivity increased or decreased.

We begin by presenting summary statistics at the municipality level. Table 8 shows that while the annualized daily traffic from a source to a different municipality increased from 72,162 vehicles in 2005 to 183,229 in 2015, overall connectivity to the national highway network declined, reflected in a negative z-score of -8,029.43. In other words, although absolute traffic volumes increased, those flows became more concentrated along certain routes, leaving many municipalities with weaker ties to the national highway system.

We illustrate the change in the network of annualized daily vehicle flows across municipalities in Fig. 10. In this network diagram, nodes represent municipalities and edges represent the connections estimated using the NC backboning algorithm. Green highlights municipalities that saw increased connectivity, yellow shows neutral changes, orange indicates moderate decline, and red marks a strong decline. This visualization makes clear how, despite overall traffic growth, many municipalities experienced a relative weakening of their ties to the national highway network.

Fig. 10 also shows how the network is strongly spatially embedded, reflecting the inherent relationship between vehicle flows and the highway infrastructure that supports them. As with the previous network figure, we omit the names of most municipalities to improve readability and avoid clutter, highlighting only a few notable cases. Several municipalities with growing connectivity are located near Mexico City, such as in Hidalgo state (Pachuca) and the Estado de México (Ecatepec, Tecámac, and Chicoloapan). Municipalities experiencing significant declines in vehicle traffic are concentrated in areas closely linked to oil theft, such as Puebla, and regions where drug cartels contest territory, such as Michoacán (e.g., Zamora and Uruapan). In Michoacán, in particular, local entrepreneurs, including the country's largest exporters of avocados and lemons, face frequent extortion by organized crime to be allowed to cross the national highways.

**Table 8**

Descriptive statistics of annualized vehicles analysis

| | Mean | Standard deviation | Observations |
|---|---|---|---|
| Annualized daily number of automobiles reaching another municipality in 2005 | 72161.940 | 95100.600 | 1,228 |
| Annualized daily number of automobiles reaching another municipality in 2015 | 183228.800 | 217881.600 | 1,228 |
| Z-score of change in highway network between year 2005 and 2015 | -8029.425 | 25595.45 | 1,226 |
| Change in homicide rate between the years 2005 and 2015 | 8.694 | 26.822 | 1,203 |
| Change in the Gini coefficient between the years 2005 and 2015 | -0.033 | 0.039 | 1,203 |
| Change in poverty between the years 2005 and 2015 | -0.214 | 0.093 | 1,203 |
| Change in nightlight per capita between the periods 2001-2005 and 2011-2015 | -0.200 | 0.209 | 1,203 |
| Average number of drug trafficking organizations by municipality in 2005 | 0.254 | 0.661 | 1,203 |
| Average number of drug trafficking organizations by municipality in 2010 | 1.002 | 1.405 | 1,203 |

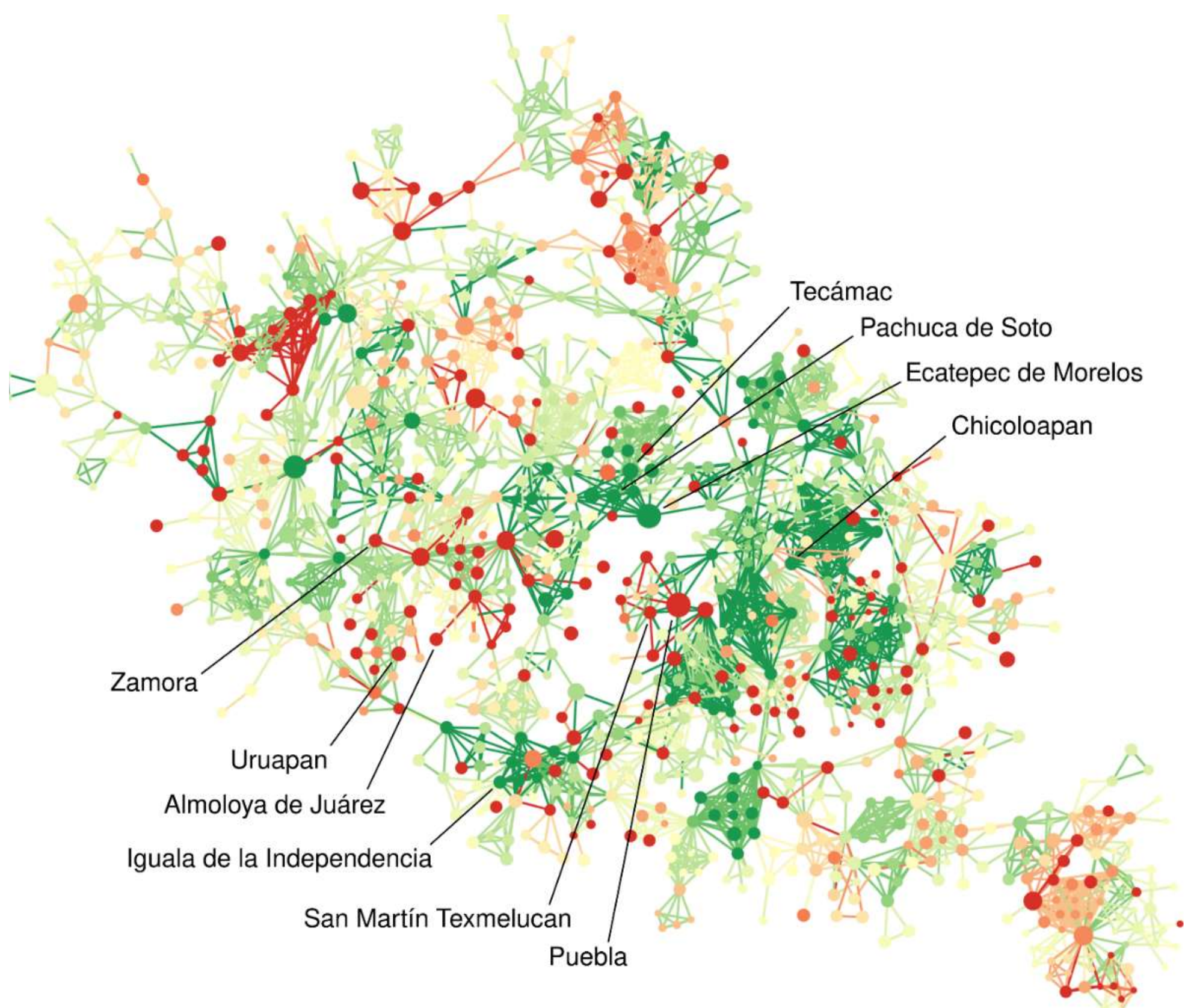

**Fig. 10.** Change in the annualized daily vehicle trip network between 2005 and 2015

To analyze the role of the changes in homicide rates between 2005 and 2015 on the changes detected in the network of vehicle trips, we next use Ordinary Least Squares (OLS). Our dependent variable is the z-scores of annualized daily vehicle flows. To complement this analysis, we include as control variables the change in municipalities' annual nighttime light-per-capita between 2001-2005 and 2011-2015, a proxy for economic activity during these periods. We also control for inequality by tracking changes in the Gini coefficient between the years 2005 and 2015, and the change in poverty rates over the same period. As mentioned earlier, the 2005 poverty rate is derived from the 2000 census, given that the 2005 mid-census did not collect household income data.

The OLS results in Table 9, columns 1 and 4, show no statistically significant association between changes in homicide rates and either the structure of the vehicles driving on the highway network (column 1) or the volume of annualized daily vehicle flows (column 4). It is important to clarify that our highway network measure does not reflect the number of highways built, but rather the degree of functional connectivity between municipalities, defined by the average number of vehicles per day that travel from a given municipality to another. While the OLS coefficients are not statistically significant, they are likely biased due to endogeneity.

Unobserved local shocks, such as declining state capacity, shifts in criminal governance, or regional stagnation, can simultaneously increase violence and suppress traffic flows. If these factors are not fully captured by observable controls, the OLS estimates will conflate the effect of homicides with broader local deterioration, biasing the coefficient. The direction of this bias depends on the nature of the omitted shocks, but in this particular case, the OLS estimates will understate the true effect of homicide rates if institutional decline or insecurity both increase violence and reduce mobility. To address this, we implement an instrumental variables strategy.

In the earlier migration analyses, we used as instruments the five-year change in oil prices and its ratio to each municipality's distance from the nearest oil pipeline. These variables plausibly captured variation in violence driven by criminal competition over fuel theft, indirectly influencing migration by increasing local insecurity. However, when applied to vehicle traffic, these instruments become problematic. Recurrent increases in fuel prices may reduce daily trips directly, by discouraging non-essential travel, shifting commuters from private vehicles to public transport, or suppressing inter-municipal visits. While we argue that nationwide fuel price changes are unlikely to directly influence long-run migration decisions, they plausibly affect short-run traffic volumes independently of violence. This violates the exclusion restriction in the context of vehicle flows. We therefore adopt a different set of instruments to identify the causal effect of violence on the daily vehicle mobility network.

To identify the causal effect of homicide rates on highway traffic, we instead exploit the variation in local violence generated by the geographic expansion and fragmentation of drug trafficking organizations (DTOs) during the early stages of Mexico's militarized drug war. Beginning in the mid-2000s, federal crackdowns on cartel strongholds led to internal splintering and spatial reorganization of DTOs, which in turn, fueled sharp increases in local homicides. This reconfiguration was driven largely by national enforcement priorities and cartel competition, rather than by local mobility patterns, making it a plausible source of exogenous variation in violence.

We construct two instruments from this context. The first is the change in the number of DTOs operating in each municipality between 2005 and 2010, based on data compiled by Coscia and Rios (2012). These authors, drawing on multiple online sources, including online newspapers and blogs, mapped the number of DTOs operating in each municipality from the 1990s until 2010. During the initial phase of Mexico's drug war, the number of DTOs was limited to a small number of areas but expanded rapidly across municipalities as militarization efforts targeted known strongholds. This spread of DTOs, as earlier research has documented (Castillo, Mejía, and Restrepo 2020), contributed to a rise in homicide rates in the country. In a second IV specification, we incorporate the number of DTOs operating in each municipality in 2005, providing a baseline prior to the militarization enforcement against criminal groups. This instrument captures the early distribution of DTO activity and its potential to predict where violence would intensify following militarization.

The exclusion restriction is credible in this context. In many municipalities, drug trafficking organizations operated for years, often decades, without significantly disrupting the lives of ordinary citizens in these areas. Their presence before the federal militarization campaign was relatively stable and did not directly interfere with daily mobility. The key change was the level of violence: as DTOs fragmented and territorial disputes escalated, so did homicide rates and associated risks to civilians. These are the mechanisms through which traffic flows were likely affected, not through DTO presence itself, but by its impact on violence. We therefore interpret our instruments as generating exogenous variation in violence that allows us to identify its causal impact on vehicle mobility.

Table 9 reports the OLS and IV estimates of the effect of homicide rates on two outcomes: the structure of the highway traffic network (columns 1–3) and the level of annualized daily vehicle flows (columns 4–5). The first-stage IV regressions are shown at the bottom of the table. As expected, both instruments (the change in the number of DTOs between 2005 and 2010, and baseline DTO presence in 2005) are positively and significantly associated with changes in homicide rates. The F-statistic for the excluded instrument in the first IV specification, columns 2 and 5, is above 20, indicating strong relevance. When both instruments are included (columns 3–6), the first-stage F-statistic falls to 11.33, still above the standard threshold. The endogeneity test also confirms the presence of endogeneity, suggesting that the IV specifications should be preferred over the OLS results as they provide a more credible estimate of the causal effect of violence on highway connectivity and changes in vehicle traffic.

**Table 9**

Change in z-standardized scores for annualized vehicles crossing municipalities between 2005 and 2015

| | (1) Highway network z-score | (2) Highway network z-score | (3) Highway network z-score | (4) Annualized daily vehicle traffic | (5) Annualized daily vehicle traffic | (6) Annualized daily vehicle traffic |
|---|---|---|---|---|---|---|
| | OLS | IV | IV | OLS | IV | IV |
| Change in the homicide rate between 2005 and 2015 | 28.17 | -678.8* | -668.2** | 82.62 | -10,588*** | -9,827*** |
| | (19.87) | (355.5) | (306.2) | (127.3) | (3,144) | (2,689) |
| Constant | -4,808*** | 1,670 | 1,574 | -76,269*** | 21,509 | 14,537 |
| | (1,735) | (3,674) | (3,436) | (9,888) | (33,891) | (31,809) |
| Controls | Yes | Yes | Yes | Yes | Yes | Yes |
| Number of municipalities | 1,203 | 1,203 | 1,203 | 1,203 | 1,203 | 1,203 |
| | | First-stage regression Change in the homicide rate | First-stage regression Change in the homicide rate | | First-stage regression Change in the homicide rate | First-stage regression Change in the homicide rate |
| Change in the number of drug trafficking organizations between 2005 and 2010 | | 2.709*** | 2.480*** | | 2.709*** | 2.480*** |
| | | (4.50) | (4.12) | | (4.50) | (4.12) |
| Number of drug trafficking organizations in 2005 | | | 2.580** | | | 2.580** |
| | | | (2.05) | | | (2.05) |
| Change in the Gini index | | -45.26** | -42.07** | | -45.26** | -42.07** |
| | | (-2.81) | (-2.60) | | (-2.81) | (-2.60) |
| Change in the poverty rate | | 14.22 | 14.25 | | 14.22 | 14.25 |
| | | (1.41) | (1.41) | | (1.41) | (1.41) |
| Change in nighttime light-per-capita | | -7.043 | -7.790 | | -7.043 | -7.790 |
| | | (-1.46) | (-1.57) | | (-1.46) | (-1.57) |
| Constant | | 6.821** | 6.298** | | 6.821** | 6.298** |
| | | (2.74) | (2.50) | | (2.74) | (2.50) |
| Number of municipalities | | 1203 | 1203 | | 1203 | 1203 |
| F test of excluded instruments: | | 20.24 | 11.33 | | 20.24 | 11.33 |
| Prob > F | | 0.00 | 0.00 | | 0.00 | 0.00 |
| Stock-Yogo weak ID test critical values: 15% maximal IV size | | 12.83 | 12.83 | | 12.83 | 12.83 |
| Hansen J statistic (overidentification test of all instruments) | | exactly identified | 0.00 | | exactly identified | 0.22 |
| Chi-sq(1) P-val | | | 0.95 | | | 0.64 |
| Endogeneity test of endogenous regressors: | | 4.82 | 7.17 | | 24.25 | 33.12 |
| Chi-sq(1) P-val | | 0.03 | 0.01 | | 0.00 | 0.00 |

Other controls include changes in the poverty rate, annual nighttime light-per-capita, and the Gini coefficient. Robust standard errors, clustered at the municipality level, are in parentheses. Significance levels *** p<0.01, ** p<0.05, * p<0.1.

The second-stage IV results (column 2) reveal a strong negative effect of homicide rates on changes in the network of vehicle trips connecting municipalities. A one-unit increase in the homicide rate results in a statistically significant decrease of 678.8 points in the z-score of the vehicle trips network. The change is slightly lower, -668.2 if looking at the second specification (column 3). This decline in connectivity suggests that increasing violence deters travel to affected areas, likely due to heightened risks of victimization, as well as migration decisions made to avoid these regions, as discussed earlier. This disruption has been reflected in reduced connectivity in the network of affected areas with the rest of the country.

We now turn to the effect of homicide rates on the volume of daily vehicle flows on highways. The second-stage IV results in Table 9, column 5, indicate that a one-unit increase in the homicide rate leads to a reduction of approximately 10,588 in the number of annualized daily vehicle trips arriving from other municipalities. The estimate is slightly smaller in column 6 when both instruments are used. These results suggest that rising homicide rates substantially reduce inter-municipal traffic volumes. The magnitude of the effect highlights how lethal violence not only alters the strength of connectivity among municipalities but also suppresses the actual flow of vehicles within the national road network.

On average, between 2005 and 2015, municipalities experienced an increase in their homicide rate of 8.69 points (Table 8). This implies that, in total, among all the 1,203 municipalities with highway connections, rises in homicide rates caused an estimated decrease of around 110 million daily vehicle trips originating from outside their boundaries (10,588 × 1,203 × 8.69).

The estimated loss of 110 million daily vehicle trips represents approximately 49% of total daily intermunicipal highway traffic the country had in 2015. It is possible that some of this decline may be due to people carpooling to share driving costs, reduce their risk of victimization, or opting for bus travel instead of private vehicles. Still, the large fall in daily vehicle flows reaching a different municipality suggests a growing disconnection of violent areas, as evidenced by shifts in migration and highway networks.

## 7. Conclusion

This paper advanced our understanding of how violence reshaped migration by affecting not only the volume of migration flows but also the structure of migration networks, an aspect largely overlooked in the existing literature. Our study showed that shifts in origin-destination pathways, such as new routes emerging and old ones disappearing, can be overlooked if the analysis focuses solely on changes in the number of migrants. By analyzing changes in Mexico's domestic and international migration network from 2005 to 2020, we demonstrated how violence has reconfigured both domestic and international migration flows and networks.

Using individual-level census data and a modified Noise-Corrected algorithm, we quantified changes in over three million bi-directional domestic migration flows among all municipalities in Mexico and their international connections. Then, in a second step, we used instrumental variables to analyze what the causal role of violence, proxied by changes in overall homicide rates, was in driving the change in migration networks observed. Our findings have broad implications for researchers studying network dynamics, the economics of violence, and stability in conflict-affected countries.

We provided four key contributions. First, we show that rising homicide rates have fundamentally reshaped Mexico's domestic migration network, displacing more people from violent areas than those arriving. While violent municipalities still attract some new residents, emigration occurs at a greater rate, highlighting violence as a significant push factor. This mirrors patterns seen in other countries where violence drives people away given the risks of victimization (Melander and Öberg 2007; Bosetti, Cattaneo, and Peri 2021), even as some movement into these areas persists (Ceriani and Verme 2018). Our network analysis adds an important layer of insight. It reveals that emigration consistently outpaced immigration in municipalities affected by violence, pushing people into a broader range of destinations and reshaping the broader domestic migration networks.

Second, earlier studies documented a substantial increase in the return of Mexican migrants from the U.S. since 2010 (Pearson 2021). Our network analysis adds nuance by showing that while the structure of the return migration network has remained largely unchanged, rising violence in Mexico has reduced the pace of return migration to violent areas. This finding is particularly relevant in the context of the current U.S. debate on deportations, and the reintegration of families in their countries of origin (Amuedo-Dorantes, Pozo, and Puttitanun 2015; Rosenblum et al. 2014). With violence still high in Mexico, our analysis suggests that many Mexicans in the U.S. have reduced incentives to return to the most violent municipalities. This insight also contributes to the field of migration studies by demonstrating that return migration, often seen as a response to economic and family reunification factors (Hazán 2014), is also sensitive to local conditions of violence.

Third, while violence has expanded non-transit emigration to the U.S., its most profound impact is on internal migration within Mexico, highlighting the scale of internal displacement driven by violence. This challenges the conventional focus on South-to-North migration and highlights the need to prioritize internal displacement in policy discussions. This is particularly important in developing countries like Mexico, where sudden armed conflict has left the country ill-prepared to even count how many families are being displaced in the short and medium term. Finally, as a robustness check, we showed that rising homicide rates have also led to a long-term decline in 110 million daily vehicle trips on highways connecting municipalities to one another in the broad national highway network, reflecting a broader impact on regional connectivity.

In conclusion, our findings suggest that violence disrupts migration flows, the origin and destiny of these flows, along with the underlying migration networks that link areas. While the study focused on Mexico, the implications extend to other conflict-prone regions, offering a framework for assessing how violence and instability reshape migration networks and economic linkages.

**References**

Adhikari, Prakash. 2013. "Conflict-Induced Displacement, Understanding the Causes of Flight." *American Journal of Political Science* 57 (1): 82–89.

Amuedo-Dorantes, Catalina, Susan Pozo, and Thitima Puttitanun. 2015. "Immigration Enforcement, Parent-Child Separations, and Intent to Remigrate by Central American Deportees." *Demography* 52 (6): 1825–51.

Arzate, Esther. 2023. "Acabar Con El Huachicol, Promesa Incumplida." *Forbes*, July 3, 2023.

Atuesta, Laura H., and Dusan Paredes. 2015. "Do Mexicans Flee from Violence? The Effects of Drug-Related Violence on Migration Decisions in Mexico." *Journal of Ethnic and Migration Studies* 42 (3): 480–502.

Basu, Sukanya, and Sarah Pearlman. 2017. "Violence and Migration: Evidence from Mexico's Drug War." *IZA Journal of Development and Migration 2017 7:1* 7 (1): 1–29.

Battiston, Giacomo, Gianmarco Daniele, Marco Le Moglie, and Paolo Pinotti. 2024. "Fuelling Organised Crime: The Mexican War on Drugs and Oil Theft." *The Economic Journal* 134 (663): 2685–2711.

Blattman, Christopher, and Edward Miguel. 2010. "Civil War." *Journal of Economic Literature* 48 (1): 3–57.

Bosetti, Valentina, Cristina Cattaneo, and Giovanni Peri. 2021. "Should They Stay or Should They Go? Climate Migrants and Local Conflicts." *Journal of Economic Geography* 21 (4): 619–51.

Bucheli, José R., Matías Fontenla, and Benjamin James Wadell. 2019. "Return Migration and Violence." *World Development* 116: 113–24.

Calderón, Gabriela, Gustavo Robles, Alberto Díaz-Cayeros, and Beatriz Magaloni. 2015. "The Beheading of Criminal Organizations and the Dynamics of Violence in Mexico:" *Journal of Conflict Resolution* 59 (8): 1455–85.

Castillo, Juan Camilo, Daniel Mejía, and Pascual Restrepo. 2020. "Scarcity without Leviathan: The Violent Effects of Cocaine Supply Shortages in the Mexican Drug War."

*Review of Economics and Statistics* 102 (2): 269–86.

Ceriani, Lidia, and Paolo Verme. 2018. "Risk Preferences and the Decision to Flee Conflict." *World Bank Policy Research Working Paper*. World Bank, Washington, DC.

Coscia, Michele, and Frank M.H. Neffke. 2017. "Network Backboning with Noisy Data." *Proceedings - International Conference on Data Engineering*, May, 425–36.

Coscia, Michele, and Viridiana Rios. 2012. "Knowing Where and How Criminal Organizations Operate Using Web Content." *CIKM*, 2012.

Daniele, Gianmarco, Marco Le Moglie, and Federico Masera. 2023. "Pains, Guns and Moves: The Effect of the U.S. Opioid Epidemic on Mexican Migration." *Journal of Development Economics* 160 (January): 102983.

Dell, Melissa. 2015. "Trafficking Networks and the Mexican Drug War." *American Economic Review* 105 (6): 1738–79.

Dittmar, Victoria. 2018. "The Mexico Crime Bosses Peña Nieto's Government Toppled." *InSight Crime*, September 24, 2018.

Elvidge, Christopher D., Kimberly Baugh, Mikhail Zhizhin, Feng Chi Hsu, and Tilottama Ghosh. 2017. "VIIRS Night-Time Lights." *International Journal of Remote Sensing* 38 (21): 5860–79.

Ferri, Pablo. 2019. "Huachicolero: 'Antes, Como Policía, Ganaba 270 Dólares Al Mes. Ahora, Con El Combustible, Puedo Sacar Hasta 50.000.'" *El País*, January 15, 2019.

Gutiérrez-Romero, Roxana. 2016. "Estimating the Impact of Mexican Drug Cartels and Drug-Related Homicides on Crime and Perceptions of Safety." *Journal of Economic Geography* 16 (4): 941–73.

———. 2024. "Drug Trafficking Fuels Violence Leading to Mass Emigration: The Case of Guatemala." *Economic Modelling* 131 (February): 106595.

Gutiérrez-Romero, Roxana, and Nayely Iturbe. 2024. "Causes and Electoral Consequences of Political Assassinations: The Role of Organized Crime in Mexico." *Political Geography* 115 (November): 103206.

Gutiérrez-Romero, Roxana, and Mónica Oviedo. 2018. "The Good, the Bad and the Ugly: The Socioeconomic Impact of Drug Cartels and Their Violence." *Journal of Economic Geography* 18 (6): 1315–38.

Hazán, Miryam. 2014. "Understanding Return Migration to Mexico: Towards a Comprehensive Policy for the Reintegration of Returning Migrants." Center for Comparative Immigration Studies Working Paper 193.

International Crisis Group. 2022. "Keeping Oil from the Fire: Tackling Mexico's Fuel Theft

Racket," 2022.

Jones, Nathan P., and John P. Sullivan. 2019. "Huachicoleros: Criminal Cartels, Fuel Theft, and Violence in Mexico." *Journal of Strategic Security* 12 (4): 1–24.

Magaloni, Beatriz, Gustavo Robles, Aila M. Matanock, Alberto Diaz-Cayeros, and Vidal Romero. 2019. "Living in Fear: The Dynamics of Extortion in Mexico's Drug War:" *Comparative Political Studies* 53 (7): 1124–74.

Melander, Erik, and Magnus Öberg. 2007. "The Threat of Violence and Forced Migration: Geographical Scope Trumps Intensity of Fighting." *Civil Wars* 9 (2): 156–73.

Michaelsen, Maren M., and Paola Salardi. 2020. "Violence, Psychological Stress and Educational Performance during the 'War on Drugs' in Mexico." *Journal of Development Economics* 143 (March): 102387.

Orozco-Aleman, Sandra, and Heriberto Gonzalez-Lozano. 2018. "Drug Violence and Migration Flows." *Journal of Human Resources* 53 (3): 717–49.

Pearson, Thomas. 2021. "U.S. Immigration Enforcement and Mexican Labor Markets." Boston University Working Paper.

Proceso. 2024. "Hacienda Quita El Subsidio a Gasolinas Magna, Premium y Diésel; Usuarios Pagarán El 100% Del IEPS." *Proceso*, July 7, 2024.

Rios Contreras, Viridiana. 2014. "The Role of Drug-Related Violence and Extortion in Promoting Migration. Unexpected Consequences of a Drug War." *Latin American Research Review* 49 (3): 199–2017.

Rosenblum, Marc R., Doris Meissner, Claire Bergeron, and Faye Hipsman. 2014. "The Deportation Dilemma. Reconciling Tough and Humane Enforcement." Washington, DC: United States.

Serrano, M. Ángeles, Marián Boguñá, and Alessandro Vespignani. 2009. "Extracting the Multiscale Backbone of Complex Weighted Networks." *Proceedings of the National Academy of Sciences of the United States of America* 106 (16): 6483–88.

Stevenson, Mark. 2017. "Gasoline Thieves Are out of Control and Deadly in Mexico." *Associated Press*, July 14, 2017.

Velásquez, Andrea. 2019. "The Economic Burden of Crime: Evidence from Mexico." *Journal of Human Resources* 55 (4): 0716-8072r2.

Wainwright, Tom. 2017. *Narconomics*. London: Penguin Random House.

Zolberg, Aristide R., Astri Suhrke, and Sergio Aguayo. 1993. *Escape from Violence : Conflict and the Refugee Crisis in the Developing World*. Oxf. U.P. (N.Y.).

## Appendix

**Table A.1** First-stage regression for Table 6 and Table 7

| | (1) | (2) | (3) | (4) |
|---|---|---|---|---|
| Second-stage IV regression's dependent variable: | Change in z-scores for emigration to non-U.S. countries | Change in z-scores for returning Mexican immigrants from non-U.S. | Change in number of emigrants to non-U.S. countries | Change in number of returning Mexican immigrants from non-U.S. |
| First-stage regression's dependent variable: | Change in the homicide rate | Change in the homicide rate | Change in the homicide rate | Change in the homicide rate |
| Change in the average price of diesel and gasoline divided by the municipality's average distance to the nearest oil pipeline | 0.108** | 0.103** | 0.108** | 0.103** |
| | (2.42) | (2.27) | (2.42) | (2.27) |
| Change in the price of gasoline in real pesos | 372.7*** | 370.5*** | 372.7*** | 370.5*** |
| | (5.36) | (5.33) | (5.36) | (5.33) |
| Change in the gini index | -50.07 | -37.22 | -50.07 | -37.22 |
| | (-0.77) | (-0.57) | (-0.77) | (-0.57) |
| Change in the poverty rate | -51.93 | -42.52 | -51.93 | -42.52 |
| | (-1.63) | (-1.30) | (-1.63) | (-1.30) |
| Change in the night-time light per capita | 590.4** | 605.4*** | 590.4** | 605.4*** |
| | (3.24) | (3.35) | (3.24) | (3.35) |
| Municipality fixed effects | Yes | Yes | Yes | Yes |
| Year fixed effects | Yes | Yes | Yes | Yes |
| Number of municipalities | 4700 | 4736 | 4700 | 4736 |

Other controls include changes in the poverty rate, annual nighttime light-per-capita, and the Gini coefficient. Robust standard errors, clustered at the municipality level, are in parentheses. Significance levels *** p<0.01, ** p<0.05, * p<0.1.